\documentclass[]{aa}  
\usepackage{CJK}
\usepackage[colorlinks=true,linkcolor=blue,citecolor=blue, urlcolor=blue]{hyperref}
\usepackage{natbib,hyperref}
\usepackage{graphicx}
\usepackage{txfonts}

{\newif\ifnotend
\notendtrue
\def\veclist{ABCDEFGHIJKLMNOPQRSTUVWXYZabcdefghijklmnopqrstuvwxyz.}
\def\top#1#2.{#1}
\def\tail#1#2.{#2.}
\loop\expandafter\xdef\csname v\expandafter\top\veclist\endcsname%
{{\noexpand\bf\expandafter\top\veclist}}
\edef\veclist{\expandafter\tail\veclist}
\if\veclist.\notendfalse\fi\ifnotend\repeat}
{\newif\ifnotend
\notendtrue
\def\callist{ABCDEFGHIJKLMNOPQRSTUVWXYZ.}
\def\top#1#2.{#1}
\def\tail#1#2.{#2.}
\loop\expandafter\xdef\csname c\expandafter\top\callist\endcsname%
{{\noexpand\cal\expandafter\top\callist}}
\edef\callist{\expandafter\tail\callist}
\if\callist.\notendfalse\fi\ifnotend\repeat}

\def\d{{\rm d}}
\def\Vc{v_{\rm c}}

\def\apjs{ApJS}
\def\apjl{ApJL}

\def\aap{A\&A}

\def\Gyr{\,\mathrm{Gyr}}

\def\kpc{\,\mathrm{kpc}}

\def\kms{\,\mathrm{km\,s}^{-1}}

\def\msun{\,{\rm M}_\odot}

\def\eg{{ e.g.,\ }}

\def\feh{\rm{[Fe/H]}}
\def\mgfe{\rm{[Mg/Fe]}}

\def\Omegab{\Omega_{\rm b}}

\newcommand*\samethanks[1][\value{footnote}]{\footnotemark[#1]}

\newcommand{\gkai}[1]{\begin{CJK*}{UTF8}{gkai}\raisebox{.1em}{(}#1\raisebox{.1em}{)}\end{CJK*}}

\renewcommand{\[}{\begin{equation}}
\renewcommand{\]}{\end{equation}}

\begin{document}
   \title{Exploring the impact of a decelerating bar on transforming bulge orbits into disc-like orbits}
   
   \titlerunning{Metal poor thin disc}
   \authorrunning{Li et al.}

   \author{Chengdong Li \gkai{李承东}\inst{1}\fnmsep\thanks{\email{chengdong.li@astro.unistra.fr;\\zhen.yuan@astro.unistra.fr}}
          \and Zhen Yuan \gkai{袁珍}\inst{1}\fnmsep\samethanks
          \and Giacomo Monari\inst{1}
          \and Nicolas F. Martin \inst{1,2}
          \and Arnaud Siebert\inst{1}
          \and Benoit Famaey\inst{1}
          \and Rimpei Chiba \inst{3}
          \and Georges Kordopatis \inst{4}
          \and Rodrigo A. Ibata \inst{1}
          \and Vanessa Hill \inst{4}
          }

   \institute{Universit\'e de Strasbourg, CNRS, Observatoire astronomique de Strasbourg, UMR 7550, F-67000 Strasbourg, France
         \and
         Max-Planck-Institut f\"{u}r Astronomie, K\"{o}nigstuhl 17, D-69117 Heidelberg, Germany
	    \and
	    Canadian Institute for Theoretical Astrophysics, University of Toronto, 60 St. George Street, Toronto, ON M5S 3H8, Canada
         \and
         Universit\'e C\^ote d'Azur, Observatoire de la C\^ote d'Azur, CNRS, Laboratoire Lagrange, Nice, France
 }

\date{Received XXX; accepted XXX}

  \abstract
   {}
   {The most metal-poor tail of the Milky Way ([Fe/H] $\leq$ $-$2.5) contains a population of stars with very prograde planar orbits, which is puzzling in both their origin and evolution. A possible scenario is that they are shepherded by the bar from the inner Galaxy, where many of the old and low-metallicity stars in the Galaxy are located.}
   {To investigate this scenario, we use test-particle simulations with an axisymmetric background potential plus a central bar model. The test particles are generated by an extended distribution function (EDF) model based on the observational constraints of bulge stars.}
   {According to the simulation results, a bar with constant pattern speed cannot help bring stars from the bulge to the solar vicinity. In contrast, when the model includes a decelerating bar, some bulge stars can gain rotation and move outwards as they are trapped in the bar's resonance regions. The resulting distribution of shepherded stars heavily depends on the present-day azimuthal angle between the bar and the Sun. The majority of the low-metallicity bulge stars driven outwards are distributed in the first and fourth quadrants of the Galaxy with respect to the Sun, and about 10$\%$ of them are within 6 kpc from us.}
   {Our experiments indicate that the decelerating bar perturbation can be a contributing process to explain part of the most metal-poor stars with prograde planar orbits seen in the Solar neighbourhood but is unlikely to be the only one.}


  \maketitle
\nolinenumbers
\section{Introduction}{\label{sec:intro}}

In the standard picture of galaxy formation, the innermost part of the Galaxy forms first probably in the same phase as the assembly of the inner halo where most of the oldest and most metal-poor stars ([Fe/H] $<$ $-$2.5) in the Galaxy are located \citep[see\eg][]{starkenburg17, el-badry18}. The Galactic old $\alpha$-rich thick disc (with [Fe/H] $\gtrsim$ $-$2.5) starts to form after the old bulge was built, and the thin disc is formed later with predominately much younger and more metal-rich stars \citep[see \eg][]{bovy12b, brook12, gerhard16, gallart19, xiang22}.

Based on the broad-brush picture described above, we would expect the very low-metallicity stars ([Fe/H] $<$ $-$2.5) in the solar neighborhood very likely to be debris from the very early assembly phase of our Galaxy. Observations show that there are more stars in the very low metallicity regime that have prograde orbits compared to those with retrograde orbits. Moreover, there is a population of prograde planar stars with strong rotation (azimuthal action $J_{\phi}$ $>$ 1000~km~s$^{-1}$ kpc, which represents the angular momentum) and small vertical motion (vertical action $J_z$ $<$ 400~km~s$^{-1}$ kpc, which quantifies oscillations away from the Galactic plane), similar to the disc stars \citep{Sestito2020, DiMatteo2020, Carollo2023, FernandezAlvar2024}. There are several possibilities of their origins under the debate \citep[see \eg][]{Sestito2020, Sestito2021, Yuan2023}. One possible scenario is that they can be transported by the bar from the inner Galaxy as suggested by \citet{Dillamore2023} using a halo-like population under a short bar perturbation. Also, galactic bars are common in disc galaxies. The dynamical interplays between bar, disk, stellar and dark matter halo are essential to galaxy evolution, but their understanding is far from complete \citep[see \eg][]{Sellwood1993,Sellwood2014}. In this work, we explore this possibility by tracing a peanut-shape bulge-like population generated in the framework of an extended distribution functions (EDFs) model under the perturbation of a long bar with different settings of pattern speeds. In order to make a comparison, we also carry out a simulation based on the pseudo stars generated via a double power law spheroidal distribution function (DF) model. 

\citet{Schonrich2009} modelled the Galaxy as a series of annuli with a chemical evolution process within which stars form from gas that they simultaneously enrich. Based on this work, \citet{Sharma2021} used the continuous disc model without distinct thin or thick disc to model the chemical evolution of the disc in the Galaxy. Similarly, \citet{Chen2023} used the galactic chemical evolution model with latest nucleosynthesis yields to model the chemical evolution with radial mixing in the Galactic disc. Both these works adopt a Schwarzschild DF $f(E_R, J_\phi, E_z)$ for the equilibrium model. \citet{Sanders2015} updated the DFs with actions $\vJ$ as arguments, and introduced the concept of EDF for the Galactic disc components, that is a density of stars in the five-dimensional space spanned by the actions $\vJ$ together with chemical abundances $\vc\equiv\big(\hbox{[Fe/H],\,[Mg/Fe]}\big)$. The EDF $F(\vc,\vJ)$ is thus a function of $\vJ$ and chemistry $\vc$. This EDF modelling method has also shown its ability in exploring the features of the stellar halo \citep{Das2016a,Das2016b}. 

The present work aims to investigate whether the bar can play a role in shaping the spatial distribution of the most metal-poor stars across the radial direction in the Galaxy. A test-particle simulation approach, with a background axisymmetric potential plus a bar perturbation, is adopted to explore the capability of the bar in transporting the stellar population with [Fe/H] $<$ $-$2.5 across the disc. The paper is organized as follows. The EDF model for generating initial pseudo stars and compiling the total potential of the Galaxy is introduced in Section~\ref{sec:model}. Then the two kinds of bar models adopted in the test-particle simulation are presented in Section~\ref{sec:bar}. The simulation results are shown in Section~\ref{sec:re&com}. In order to interpret the simulation results, observational data from $Gaia$ Radial Velocity Sample \citep[RVS;][]{dr3rvs} with the Pristine DR1 \citep{martin2023} are selected and the comparison with simulations are shown in Section~\ref{sec:com}. 
Finally, Section~\ref{sec:con} summarises the paper and indicates directions for future work.

\section{Extended Distribution Functions modelling and simulation scheme}{\label{sec:model}}

The origin and evolution of the most metal-poor stars population with prograde planar orbits is still in debate. One of the  possible scenario is that those stars were born the inner Galaxy and  brought outwards by the bar. In order to investigate this scenario, we carry out test-particle simulations with an axi\-symmetric background potential plus a central bar model in this work. Here we introduce the model used to generate pseudo stars and the potential of the bar. 

The Galaxy's model used in this work includes two parts, an axisymmetric background and a central bar. The scheme of the test-particle approach generates the pseudo stars based on the axisymmetric DFs, then integrates the pseudo stars' orbits in a potential including the bar. 
It should be noticed that no selection function (SF) is adopted in our analysis since the SF of the spectroscopic data used in this work is too complicated to model as they come from various observations. Nonetheless, our goal in this paper is to find a possible origin for these thin-disc like stars in the solar vicinity, instead of quantitatively fit our models to the observations. 

\subsection{The axisymmetric model of the Galaxy}{\label{sec:galaxy}}

In equilibrium, the DF can be taken as a function of the three action variables $(J_r,J_z,J_\phi)$, which are isolating integrals of the motion \citep{Binney2008}, where $J_r$ quantifies the amplitude of radial excursions. Following the modelling work in \citet{Binney2023}, an equilibrium Galaxy model can be defined by a superposition of similar DFs, including the dark halo, the stellar halo, the disc-like bulge and four disc components: a young disc, a middle-aged disc and an old thin disc plus a high-$\alpha$ disc. The gravitational potential that these DFs jointly generate can then be found iteratively by the method introduced by \citet{Binney2014} using the {\it AGAMA} package \citep{Vasiliev2019}. The DF of each Galactic component is a specified function $f(\vJ)$ of the action integrals. In this model, the disc components have exponential DFs, which is physically reasonable throughout action space compared to quasi-isothermal DFs. A double power law DF is used to model the stellar and dark halo. We do not have much observational data via the centre of the Galaxy, and obviously the central part of the Galaxy is non-axisymmetric. As a consequence, and to cover different possibilities, we use two different kinds of bulge models in our work, a truncated exponential DF which is introduced in Section~\ref{sec:trun} and a double power law DF which is introduced in Section~\ref{sec:sph}. The total gravitational potential of the Galaxy is then derived from the DFs. The details of the descriptions of this Milky Way axisymmetric model can be found in \citet[]{Binney2023,Binney2024}. The circular velocity generated from this axisymmetric potential is shown in the first panel of Figure~\ref{fig:df}. The black dots denote the circular velocity measurements from Classical Cepheids inside 10 $\kpc$ from the Galactic centre \citep{Ablimit2020}. 
Following what we proposed in \citet{Li2022b}, a taper-exponential model is adopted for the young thin disc. The taper term in the model is a result of the sharp decline of the surface density of the gas inside the giant molecular ring, where the young disc stars have formed recently. Based on such a model, the disc components have a maximum circular velocity at $6-8 \kpc$ away from the centre as it's shown in the upper panel of Figure~\ref{fig:df}. 

The EDF model in this work is based on the method introduced in \citet{Binney2024}, in which the EDF $F(\vc,\vJ)$ has $\vJ$ and chemistry $\vc\equiv\big(\hbox{[Fe/H],\,[Mg/Fe]}\big)$ as its arguments. Because of the product rule, the EDF $F(\vc,\vJ)$ is found by multiplying the DF $f(\vJ)$ by the probability density $P(\vc|\vJ)$ which is the probability that a star with actions $\vJ$ has the chemistry $\vc$, i.e. $F(\vc,\vJ)\,=\,f(\vJ)P(\vc|\vJ)$. 
The detailed EDF bulge model will then be introduced in Sections~\ref{sec:df} and \ref{sec:edf}. 

\subsection{DF of the bulge}{\label{sec:df}}

\subsubsection{Truncated exponential DF}{\label{sec:trun}}

Unlike the spheroidal bulge DF model used in \citet[]{Li2022b} and \citet[]{Binney2023}, this work follows \citet{Binney2024}, which uses a truncated exponential DF model for the bulge as:
\begin{equation}
    \label{eq:bulgeDF}
f(\vJ)=f_\phi(J_\phi)f_r(J_r,J_\phi)f_z(J_z,J_\phi)f_{\rm{ext}}(J_\phi).
\end{equation}
In this model, the function $f_r$ controls the velocity dispersions $\sigma_R$ and $\sigma_\phi$ near the Galactic plane. The function $f_z$ controls both the thickness of the disc and the velocity dispersion $\sigma_z$. The factor $f_\phi$ generates a roughly exponentially declining surface density $\Sigma(R)\simeq\exp(-R/R_\d)$. The factor $f_{\rm ext}$ truncates the disc at some outer radius. The detailed description of this model can be found in \citet{Binney2024}. 

There are two reasons for converting the bulge to a disc-like component. One is that the observational data \citep{Apogee2022} show that the bulge is almost entirely confined to $J_\phi>0$ \citep{Binney2024}, which can naturally be represented by a disc-like model with prograde orbits. Secondly, a weak net rotation has been detected from RR Lyrae samples within $\sim$ 3 kpc from the Galactic centre \citep[]{Wegg2019,Li2022a} and the metal-poor bulge sample from PIGS \citep{Arentsen2020} which is probably the sign of a boxy/peanut bulge in the inner disc \citep[see also][]{Shen2010}. Meanwhile, the dark matter halo can also show a net rotation due to the dynamical friction with the bar \citep[]{Chiba2022}. In the design of our experiment, we choose a disc-like bulge model that can produce a net rotation in the inner part of the Galaxy to approximately simulate the boxy/peanut bulge, which a spheroidal DF model cannot represent. We model the bulge as a truncated exponential with large velocity dispersions and small radial extensions. 

\begin{figure}
	\includegraphics[width=\columnwidth]{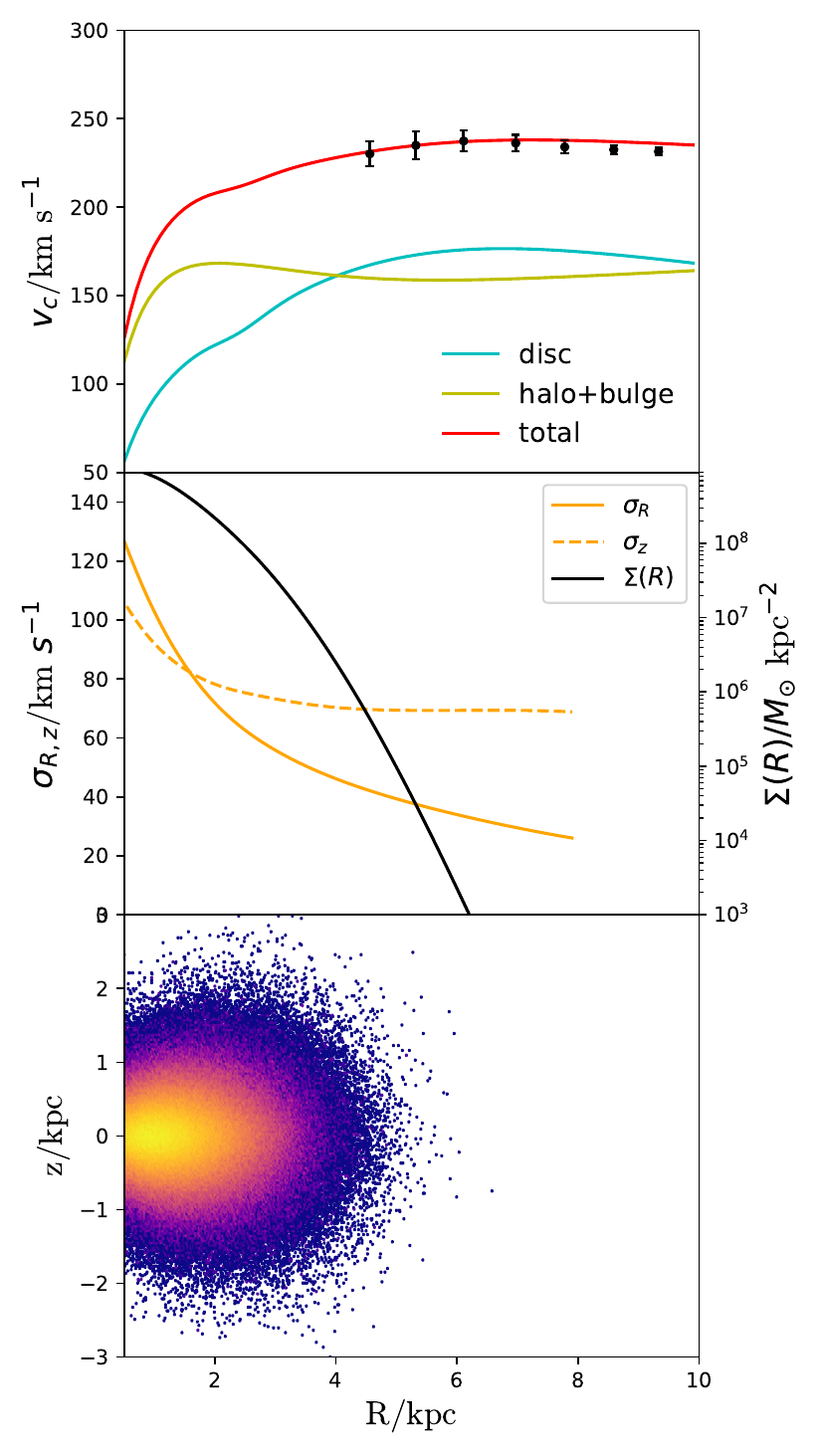}
	\vspace{-0.3cm}
    \caption{The circular velocity within 10 $\kpc$ from the Galactic centre is shown in the upper panel, in which the yellow curve represents the sum from the bulge and halo. The red curve denotes all four disc components. The black dots are from \citet{Ablimit2020} with the measurement from classical Cepheids. The velocity dispersion and surface density of the bulge are shown in the middle panel. The radial velocity dispersion $\sigma_R$ drops much rapidly relative to the vertical velocity dispersion $\sigma_z$. The bottom panel shows the spatial distribution of the pseudo stars in the ($R$, $z$) space for the bulge sample. The bulge is truncated within 6 kpc from the centre and has a peanut-like instead of a sphere shape.}
    \label{fig:df}
    \vspace{-0.3cm}
\end{figure}

The velocity dispersion and surface density of the bulge according to the parameters listed in Table~\ref{tab:df} is shown in the middle panel of Figure~\ref{fig:df}. The bottom panel shows the spatial distribution of the bulge in the ($R$, $z$) space, which is truncated at 6 kpc from the centre and has a peanut-like instead of a sphere shape.

\subsubsection{Double power law DF}{\label{sec:sph}}
In order to make a comparison, we carry out another simulation with a spheroidal DF for the bulge to test the ability of a decelerating bar to bring outwards the low metallicity stars. The perturbation and orbital integration schemes are the same with the truncated DF model. A double-power law model introduced in \citet{Li2022a} is used to generate the pseudo stars with the parameters shown in Table~\ref{tab:df_sph}. In this case, the bulge stars do not have chemical information as those boxy/peanut bulge stars generated from the EDF model mentioned above.

The double-power law distribution function for the bulge is:
\begin{equation}
    f~=~\frac{M}{(2\pi J_{0})^{3}}\frac{(1+[J_{0}/h_{J}]^{\gamma})^{\alpha/\gamma}}
    {(1+[J_{0}/g_{J}]^{\gamma})^{\beta/\gamma}}e^{-(g_{J}/J_{cut})^{\delta}},
    \label{eq:db}
\end{equation}

where M is the normalization parameter relative to the mass of the bulge. $\alpha$ and $\beta$ represent the inner and outer slopes in the phase space. $\gamma$ determines the steepness of the transition between the two parts. We keep $\gamma=1$ in this work. $J_{cut}$ is the cut-off action which makes the value of the DF drop at large radii and ${\delta}$ is the steepness for that drop in order to keep the total mass always finite in the model. $h(\mathbf{J})$ and $g(\mathbf{J})$ are defined as:
\begin{equation}
    \begin{aligned}
    &h(\mathbf{J}) = (3-h_{\phi}-h_{z})\,J_{r}\,+\,0.7(h_{z}\,J_{z}+h_{\phi}\,
    \left|J_{\phi}\right|),\\
    &g(\mathbf{J}) = (3-g_{\phi}-g_{z})\,J_{r}\,+\,0.7(g_{z}\,J_{z}+g_{\phi}\,
    \left|J_{\phi}\right|),
    \end{aligned}
    \label{eq:action}
\end{equation}
where $h_{\phi}$ and $h_{z}$ determine the velocity anisotropy at small radii and $g_{\phi}$ and $g_{z}$ determine the velocity anisotropy at large radii.

\subsection{Chemical model of the bulge}{\label{sec:edf}}

The probability that a star with actions $\vJ$ has the chemistry $\vc$ can be written as 

\begin{equation}
  \label{eq:Gauss}
P(\vc|\vJ)\,=\,\frac{\sqrt{\det(\vK)}}{{2\pi}}\, \exp\left(\frac{1}{2}(\vc-\vc_\vJ)^T\cdot\vK\cdot(\vc-\vc_\vJ)\right),
\end{equation}

where $\vK$ is the $2\times2$ covariance Matrix, and $\vc_\vJ$ depends linearly on $\vJ$ as:
\begin{equation}
  \label{eq:c}
\vc_\vJ\,=\,\vc_0\,+\,\vC(\vJ-\vJ_0).
\end{equation}
$\vc_0$ contains two specific initial values for  [Fe/H] and [Mg/Fe], and $\vC$ is a $2\times3$ matrix which connects the action $\vJ$ and chemistry $\vc$. This probability can be approximated by a Gaussian distribution in $\vc$ with mean and dispersion depending on $\vJ$ for both [Fe/H] and [Mg/Fe] according to the model by \citet{Binney2024}. 

\begin{figure}
	\includegraphics[width=\columnwidth]{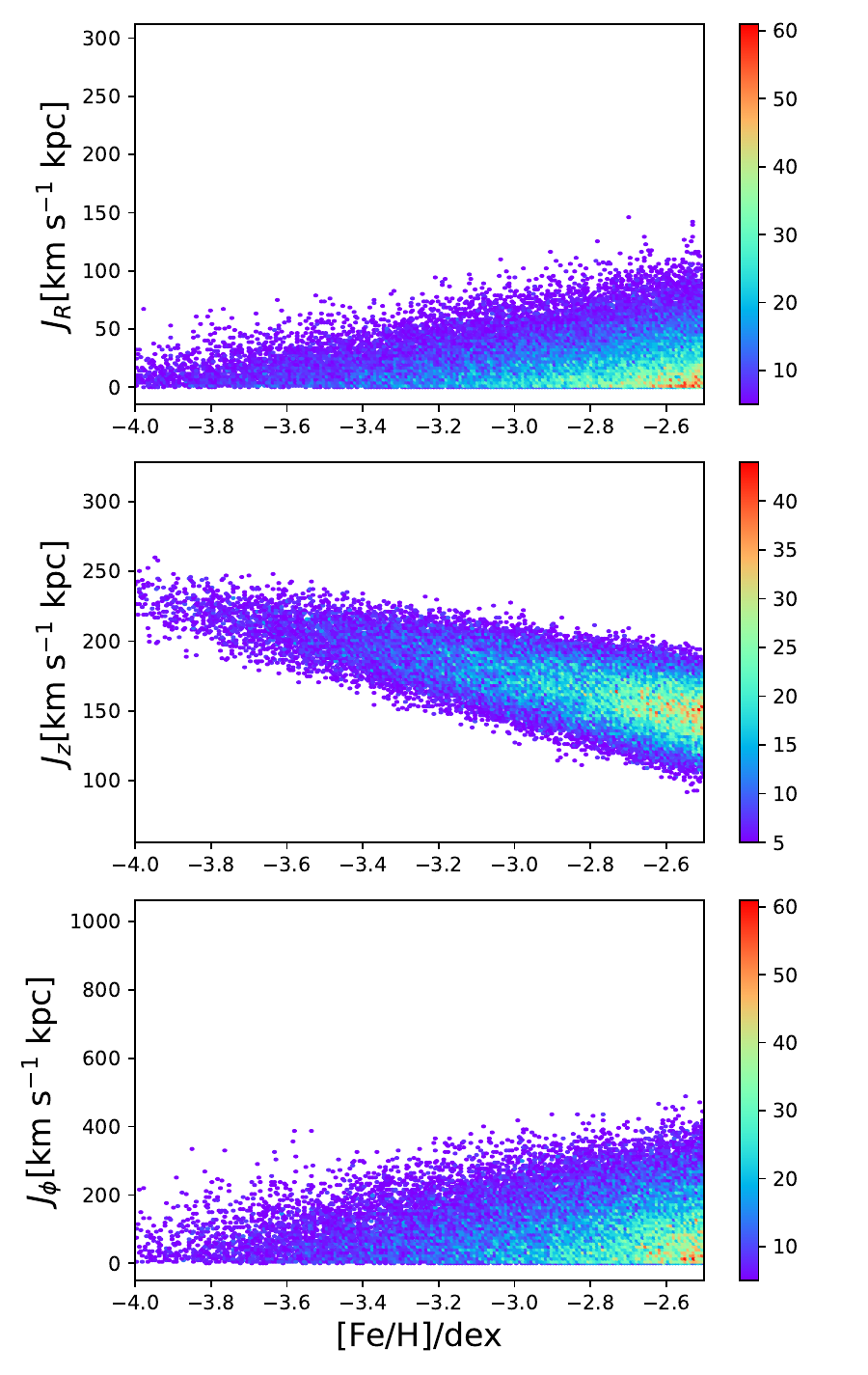}
	\vspace{-0.3cm}
    \caption{The chemical model of the bulge used in this work showing the relation between $\feh$ and actions. The three panels are colour-coded in number density for $J_{R}$, $J_{z}$, and $J_{\phi}$ respectively. As we only focus on the metal poor tail of the bulge, the $x-\rm{axis}$ is truncated at $\rm{[Fe/H]}=-2.5$ in this plot.}
    \label{fig:edf}
    \vspace{-0.3cm}
\end{figure}

Figure~\ref{fig:edf} shows the chemodynamical structure of the bulge by showing the relation between $\feh$ and the three actions. The three panels are colour-coded in number density for $J_{R}$, $J_{z}$, and $J_{\phi}$ respectively. The bulge itself is modelled by a relatively young metal rich component and most of our focus is on its old and very metal-poor parts in the simulations. The distribution extends to lower metallicity showing that the dominant dependence of chemistry on action is with $J_{z}$, with smaller $J_{z}$ for metal rich stars and larger $J_{z}$ for more metal poor stars. 

The parameters of the bulge EDF used in this work are listed in Table~\ref{tab:chem}. The parameters in this chemical model are fitted by the observation of current bulge stars \citep{Binney2024}, with some small modifications in this work. We made two modifications in the parameters. First, the value of $J_{\rm ext}$ is lowered compared to the best fit value in \citet{Binney2024} to represent a smaller bulge 5 $\Gyr$ ago at the early epoch of the bar formation. Secondly, the matrix  $\vC$ is adjusted to show a stellar population deficient in metallicity 5 $\Gyr$ ago relative to current bulge population. We cannot represent the true chemical composition of the bulge 5 $\Gyr$ ago, especially for $\mgfe$ because the current data is not enough to constrain models in the VMP region. But, since we are only interested in finding a possible origin of the metal poor thin-disc like stars, instead of tracing the $\alpha$-element enhancement history in the central part of the Galaxy, it is reasonable to use these parameters for sampling the metallicity distribution of the bulge stars 5 $\Gyr$ ago.  

\begin{table*}
	\centering
	\vspace{0.4cm}
	\caption{The parameters used for the DF model of the bulge. The units of mass and actions are $10^{10}\msun$ and $\kms\kpc$ respectively. The values listed for DF model are mainly taken from the best fit values in \citet{Binney2024} but with some differences.}
	\label{tab:df}
	\begin{tabular}{lccccccccc} 
		\hline
		$M$	&$J_{\phi0}$	&$J_{r0}$	&$J_{z0}$	&$J_{\rm ext}$	&$D_{\rm ext}$	&$p_r$	&$p_z$	&$J_{\rm v0}$	&$J_{\rm d0}$ \\ 
 \hline 
	
		$1.27$	&$127.5$	&$122.2$	&$34.21$	&$400$	&$200$	&$0.82$	&$-0.13$	&$150$	&$20$ \\
		\hline
	\end{tabular}
	\vspace{-0.4cm}
\end{table*}

\begin{table*}
	\centering
	\vspace{0.4cm}
	\caption{Parameters of the chemical models of the bulge. The units of $\theta$ are degrees while $x_0$, $y_0$,
$\sigma_x,\sigma_y$ are given in dex. The values quoted for the gradient matrices $\vC$ are in dex per $\hbox{Mpc}\kms$. The values listed for chemical models are mainly taken from the best fit values in \citet{Binney2024} but with some differences.}
	\label{tab:chem}
	\begin{tabular}{lccccccccccc} 
		\hline
		 	  $\theta$	 & $x_0$	 & $y_0$	 & $\sigma_x$	 & $\sigma_y$	& $C_{1,J_r}$	& $C_{1,J_z}$	& $C_{1,J_\phi}$	& $C_{2,J_r}$ 	& $C_{2,J_z}$ 	& $C_{2,J_\phi}$\\ 
 \hline 
	
			$-6$	&$0.409$	&$0.0598$	&$0.383$	&$0.0879$	&$-1.64$	&$-17.8$	&$-0.0464$	&$2.01$	&$6.39$	&$0.0204$\\
		\hline
	\end{tabular}
	\vspace{-0.4cm}
\end{table*}

\begin{table*}
	\centering
	\vspace{0.4cm}
	\caption{The parameters used for the spheroidal DF model of the bulge. The units of mass and actions are $10^{8}\msun$ and $\kms\kpc$ respectively. $\alpha$ and $\beta$ represent the inner and outer slopes in the double power law model. $J_{\rm cut}$ denotes the outer cut of the bulge and $\delta$ defines the sharpness of the downturn. $J_0$ is the break action. $h_{\phi}$ and $h_{z}$ determine the velocity anisotropy in the smaller radii and $g_{\phi}$ and $g_{z}$ determine the velocity anisotropy in the larger radii. It should be noticed that the values listed in this table is not a numerically fitted value for the Galaxy.}
	\label{tab:df_sph}
	\begin{tabular}{lcccccccccc} 
		\hline
		$M$		&$\alpha$	&$\beta$	&$J_{\rm cut}$	&$J_{0}$	&$h_{\phi}$	&$h_{z}$	&$g_{\phi}$	&$g_{z}$  & $\delta$\\ 
 \hline 
	
		$1.5$		&$0.85$	&$1.8$	&$280$	&$6$	&$1$	&$1$	&$1$	&$1$  &$2$\\
		\hline
	\end{tabular}
	\vspace{-0.4cm}
\end{table*}

\subsection{The bar models used in this work}{\label{sec:bar}}

We use two bar models in our simulations to investigate which is a better scenario for the Galaxy taking into account the observational constraints. 

First, a steadily rotating bar, with a given pattern speed $\Omega_{\rm{b}}$, is modelled following \citet{Chiba2022} as
\begin{equation}
    \Phi_{\rm{b}}(r,\theta,\phi,t)\,=\,\Phi_{\rm{br}}(r)\sin^2{\theta}\cos{m(\phi-\Omegab t)},
    \label{eq:phib}
\end{equation}
where $(r,\theta,\phi)$ are the spherical coordinates. We only include the quadrupole term, $m=2$, in this work. 

The radial dependence of the bar potential $\Phi_{\rm{br}}(r)$ is
\begin{equation}
    \Phi_{\rm{br}}(r)\,=\,-\frac{A\,\Vc^{2}}{2}\,\bigg(\frac{r}{r_{\rm{CR}}}\bigg)^{2}\,\bigg(\frac{b+1}{b+r/r_{\rm{CR}}}\bigg)^{5},
    \label{eq:phibr}
\end{equation}
where $A$ quantifies the strength of the bar and $\Vc$ is the value of circular velocity in the solar vicinity, i.e. $\Vc\,=\,235\,\kms$. The parameter $b$ describes the ratio between the bar scale length and the value of the co-rotation radius $r_{\rm{CR}}$. The parameters used in this work are given by \citet{Chiba2022}, with $A\,=\,0.02$ and $b\,=\,0.28$. 

The pattern speed of the bar is $\Omegab\,=\,-35\,\kms\,\kpc^{-1}$ \citep{Binney2020,Chiba2021b}. 
In the right-handed coordinate system used in this work, the pattern speed of the bar is negative and the bar rotates clockwise as the disc does. The parameters are all summarized in Table~\ref{tab:bar}. 
\begin{table}
	\centering
	\vspace{0.4cm}
	\caption{The parameters used for the bar with steady rotation. $\Omega$ is in $\kms\,\kpc^{-1}$, and $\Vc$ in $\kms$. $r_{\rm{CR}}$ is in $\kpc$. $\phi_{\rm{b}}$ denotes the initial phase angles of the bar.}
	\label{tab:bar}
	\begin{tabular}{lccccccc} 
		\hline
		Bar  & $\Omegab$ & $A$ & $\Vc$ & $b$ & $r_{\rm{CR}}$  & $\phi_{\rm{b}}$\\
	
		Values & -35 & 0.02 & 235 & 0.28 & 6.7 & $168^{\circ}$\\
		\hline
	\end{tabular}
	\vspace{-0.4cm}
\end{table}

Then, we also adopt a decelerating bar model similar to that used in \citet{Li2023}, which is based on the model by \citet{Portail2017} using a Made-to-Measure method and updated parameters by \citet{Sormani2022}. The difference with that work is that we impose the decelerating bar to end at a phase angle $\phi\,=\,28^{\circ}$  relative to the Sun, which can simulate the current morphology of the Galaxy \citep{Wegg2015}. In this work, the pattern speed of the bar drops from $\Omegab\,=\,-84\,\kms\,\kpc^{-1}$ initially to $\Omegab\,=\,-30\,\kms\,\kpc^{-1}$ at the present time. 

\begin{figure}
	\includegraphics[width=\columnwidth]{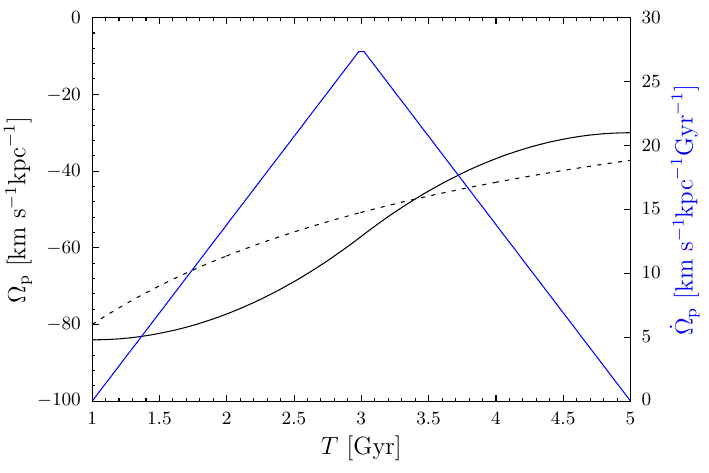}
	\vspace{-0.3cm}
    \caption{The pattern speed as a function of time is shown in solid black line. The blue curve shows the derivative of the pattern speed. For a comparison, the model used in \citet{Chiba2021a} is shown in black dashed curve.}
    \label{fig:wp}
    \vspace{-0.3cm}
\end{figure}

In this work, the pattern speed of the bar drops from $\Omegab\,=\,-84\,\kms\,\kpc^{-1}$ initially to $\Omegab\,=\,-30\,\kms\,\kpc^{-1}$ over 4Gyr. Figure~\ref{fig:wp} shows the pattern speed as a function of time (solid black) along with its derivative (blue), i.e. the slowing rate. The bar is introduced at $T=1\Gyr$ and we increase the slowing rate linearly in time until $T=3\Gyr$ where we reduce the slowing down such that $\dot{\Omega}_\textrm{p}$ is null at $T=5\Gyr$. The resulting pattern speed varies smoothly from $\Omegab\,=\,-84\,\kms\,\kpc^{-1}$ with largest variation occurring at $T=3\Gyr$. For comparison we also plot in dashed black the slowing bar model of \citet{Chiba2021a} which is constrained by the Gaia data. Although our model has a larger maximum slowing rate, the two models match reasonably well. In Appendix~\ref{sec:spp}, we report the value of the dimensionless \textit{speed parameter} of our slowing bar model \citep{Tremaine1984,Chiba2023}. This parameter is lower than one, confirming that our slowing bar model evolves in the realistic \textit{slow} regime where a substantial amount of stars is expected to be trapped in the main bar resonances.

Also, the mass and radial profile of the bar are evolving continuously during the simulation, reaching to 2.5 and 1.5 times of the initial conditions respectively to roughly simulate the growth of the bar. Note that this change in amplitude of the potential will affect the width of trapped phase-space regions, but that the variation in angular momentum for each individual star, which is dominated by churning at corotation, is regulated by the variation of the pattern speed, because stars need to be carried around by a changing corotation to change significantly their guiding radii (without settling on more eccentric orbits).

It should also be noted that, throughout this paper, we neglect the self-gravity of the disc, which is the main caveat of test-particles simulations. Following \citet{Weinberg1989}, it is known that the resonant structure can be strengthened via self-gravitating processes \citep[see also the discussions in][]{Dootson2022,Chiba2023}. This means that self-gravity will play a role in reshaping the diffusion of the disc but this is out of the scope of this work, which remains a preliminary qualitative analysis rather than a quantitative one.

The orbital integration routine of \texttt{AGAMA} is adopted to compute the trajectories of all the mock stars for 5 $\Gyr$ in total and the trajectories are stored every 0.05 $\Gyr$. In the first 1 $\Gyr$, no bar is included and pseudo stars only evolve in the axisymmetric background potential. The bar is added to the total potential at $T = 1.0 \Gyr$ with initial radial profile and a pattern speed of $\Omegab\,=\,-84\,\kms\,\kpc^{-1}$. Then the bar starts to decelerate and grow in mass and radial direction, reaching a pattern speed of $\Omegab\,=\,-30\,\kms\,\kpc^{-1}$ at the end of the simulation.

\begin{figure}
	\vspace{-0.5cm}
        \includegraphics[width=\columnwidth]{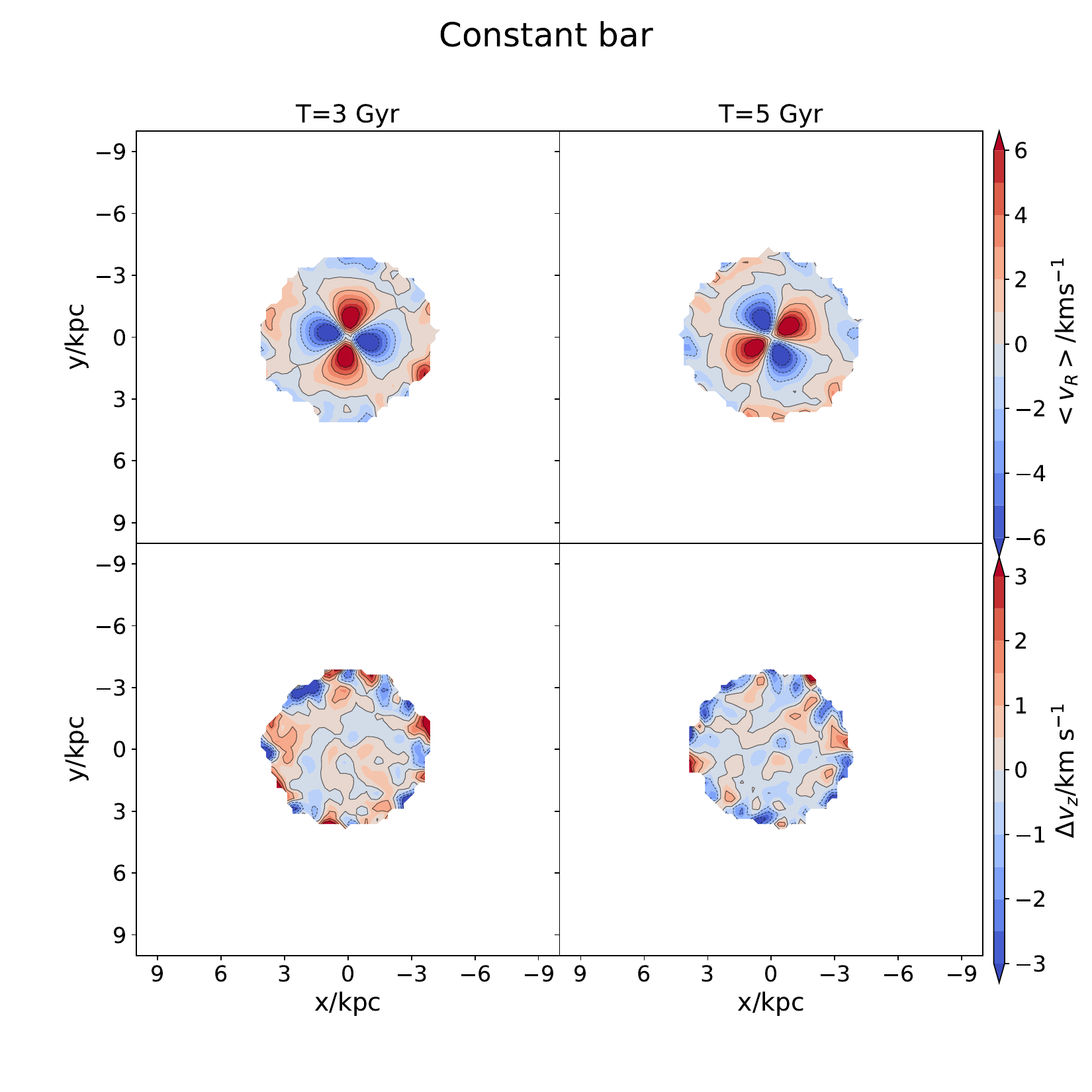}\\
	\vspace{-0.8cm}
        \includegraphics[width=\columnwidth]{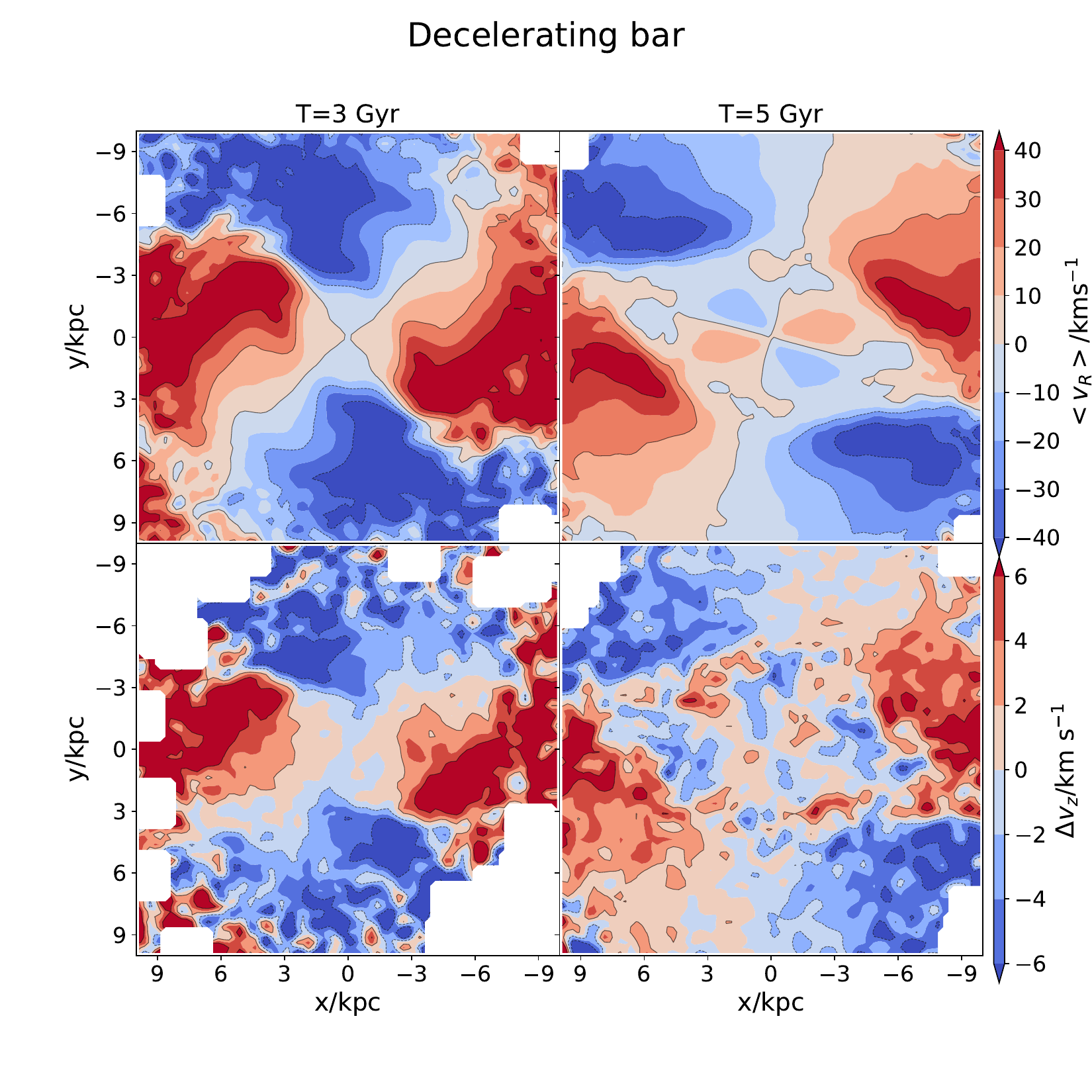}
    \vspace{-0.3cm}
    \caption{Contours of $\langle v_R \rangle$ and $\Delta_{v_z}$ for constant bar and decelerating bar respectively at $\rm{T}\,=\,3.0\,\Gyr$ and $\rm{T}\,=\,5.0\,\Gyr$. The upper panel shows the contour maps for constant bar model, which is Gaussian smoothed with a filter width of $4\,\Delta\mbox{pix}$, with $\Delta \mbox{pix}=0.25\times0.25\,(\kpc\times\kpc)$. The lower panel shows the same contour map for decelerating bar model, but is Gaussian smoothed with a filter width of $8\,\Delta\mbox{pix}$. The difference of choice in the filter width is due to the different spatial distribution of the test particles in the two simulations. The $\langle v_R \rangle$ values are several times larger than that of the constant pattern speed bar model, which suggests that the decelerating bar can generate very strong quadruple features for all the pseudo stars including those with high orbital inclinations.}
    \label{fig:cont}
    \vspace{-0.55cm}
\end{figure}

In Figure~\ref{fig:cont}, we show the face-on contour maps of $\langle v_R \rangle$ and $\Delta v_z$ for all the test particles at $\rm{T}\,=\,3.0\,\Gyr$ and $\rm{T}\,=\,5.0\,\Gyr$ for the steadily rotating and decelerating bar respectively. The value of $\Delta v_z$ is the difference between $\langle v_z \rangle$ for $z > 0$ and for $z < 0$. The upper panels correspond to the $\langle v_R \rangle$ and $\Delta v_z$ distributions for the constant pattern speed bar model at $\rm{T}\,=\,3.0\,\Gyr$ (left) and $\rm{T}\,=\,5.0\,\Gyr$ (right). The $\langle v_R \rangle$ plots clearly shows the signal of a quadruple feature in the centre of the Galaxy, whose strength differs very little between two snapshots. This is a consequence of the pseudo-stars reaching a near-equilibrium in the bar's rotating frame. On the other hand, the distribution of $\Delta v_z$ does not show any signal of a quadruple feature even in the very centre of the Galaxy. This is due to the weak bar we choose in this model, which is not strong enough to disturb the orbits with high inclination with respect to the plane. Since the initial model of the bulge is disc like with large velocity dispersions, the orbits with high inclination should make up a large fraction of the population of pseudo stars.

\begin{figure}
	\vspace{-0.8cm}
        \includegraphics[width=\columnwidth]{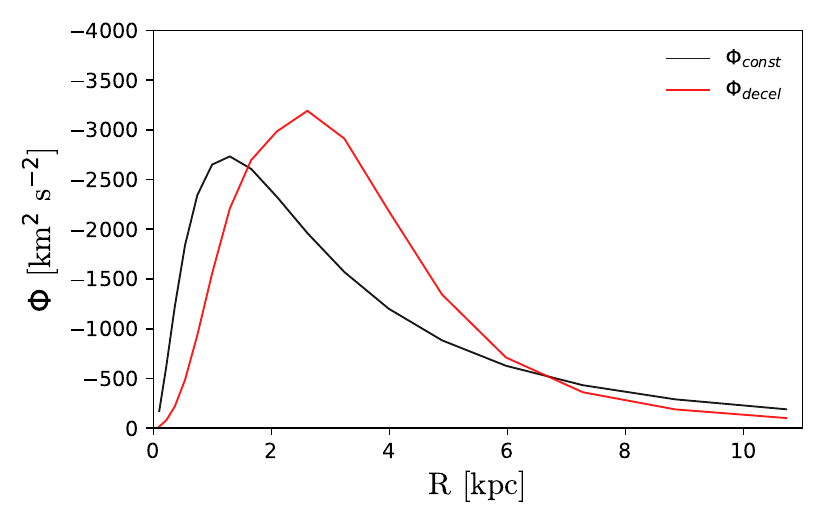}
    \vspace{-0.3cm}
    \caption{$m=2$ Fourier terms as a function of distance  at $z=0$ for both the constant rotating bar and the slowing down bar models. For the slowing down bar, the mass of the bar increases with a factor of 2.5 during the evolution of the bar while the mass of the bar keeps stable for the constant rotating bar.}
    \label{fig:potboth}
    \vspace{-0.55cm}
\end{figure}

The lower panels of Figure~\ref{fig:cont} shows the distributions of $\langle v_R \rangle$ and $\Delta v_z$ for decelerating bar model for $\rm{T}\,=\,3.0\,\Gyr$ and $\rm{T}\,=\,5.0\,\Gyr$ respectively. The pattern speed of the bar starts approximately with $\Omegab\,=\,-57\,\kms\,\kpc^{-1}$ at $\rm{T}\,=\,3.0\,\Gyr$ and keeps decreasing to $\Omegab\,=\,-30\,\kms\,\kpc^{-1}$ at $\rm{T}\,=\,5.0\,\Gyr$. The quadruple feature at $\rm{T}\,=\,3.0\,\Gyr$ is very prominent in both $\langle v_R \rangle$ and $\Delta v_z$ plots. The $\langle v_R \rangle$ values are several times larger than those of the constant pattern speed bar model, which suggests that our decelerating bar model can generate very strong quadruple features for all the pseudo stars including those with high orbital inclinations. The strong quadruple feature does not result from the deceleration but results from the forces generated by the bar, which are quite different between the two bar models. The $m=2$ Fourier terms are shown in Figure~\ref{fig:potboth} for both the constant rotating bar and the slowing down bar. The potential of the $m=2$ terms for both models have similar outputs but are only about 2$\%$ of the background axisymmetric potential of the Galaxy. 
We notice also clear quadruple features at $\rm{T}\,=\,5.0\,\Gyr$ in $\langle v_R \rangle$ and $\Delta v_z$ between $R\,\sim\,5\,\kpc$ to $R\,\sim\,10\,\kpc$ that are related to the co-rotation resonance of the bar. This shows that the strong bar plays an important role in bringing the bulge stars to a disc region with resonance trapped orbits. Meanwhile, the small quadruple features inside $R\,\sim\,4\,\kpc$ may relate to the inner Lindblad resonance. The two families of orbits have a distinct separation between each other. 

The evolution of the pseudo stars under the influence of this decelerating bar is very clear in Figure~\ref{fig:cont}. While at $\rm{T}\,=\,3.0\,\Gyr$, only one quadrupole feature is seen throughout the $x-y$ plane, at $\rm{T}\,=\,5.0\,\Gyr$, a clear separation is seen between the inner and outer quadruple features, which shows the evolution process of the quadrupole structures for the pseudo stars by the slowing down mechanism of the bar's pattern speed. With the continuous decelerating of the bar's pattern speed, the shape of the quadrupole structures continues to change without reaching a steady state in this simulation.

\section{Simulation results and comparisons with observations}{\label{sec:re&com}}
\subsection{Model Predictions}{\label{sec:pred}}

\begin{figure*}
	\includegraphics[scale=0.60]{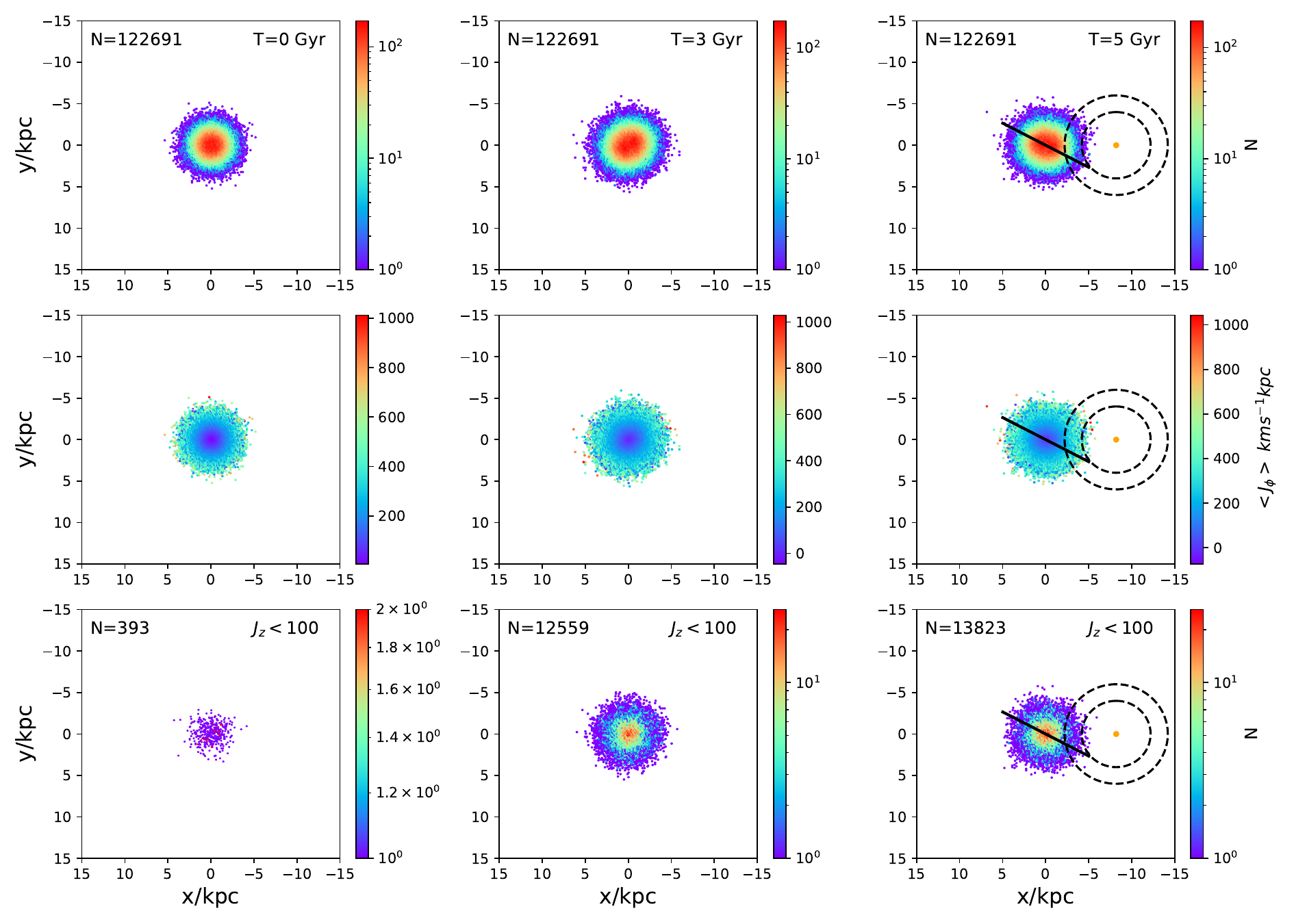}
	\vspace{-0.3cm}
    \caption{The very low-metallicity stars ([Fe/H] $\leq$ $-$2.5) for the boxy/peanut bulge model in the constant rotating bar at $\rm{T}=3\,\Gyr$ and $\rm{T}=5\,\Gyr$ middle and right respectively. The central row is colour-coded by $J_{\phi}$. The bottom row shows the low-metallicity particles with $J_{z}<100\,\kms\,\kpc$. The orange dot in the right panel marks the position of the Sun, and the two dashed curves denote the distances at 4 and 6 $\kpc$ from the Sun. The thick line denotes the direction of major axis of the bar at $\rm{T}=5\,\Gyr$. Within this constant bar model, no test particles can exceed a value of $1000 \kms\kpc$ for the value of $J_\phi$.}
    \label{fig:const_vmp}
    \vspace{-0.3cm}
\end{figure*}

In this section, we try to interpret the simulation's results for the boxy/peanut bulge model. Since we are interested in the origin and evolution of the low-metallicity stars in solar neighbourhood, we shall then focus on the test particles with [Fe/H] $\leq$ $-$2.5 from the simulations.  The left panel in Figure~\ref{fig:const_vmp} shows the initial condition of the low-metallicity particles selected according to the criteria introduced in Section~\ref{sec:intro} with $-4<\feh<-2.5$. Their density distributions at $\rm{T}=0\,\Gyr$ are shown in the top row. As the number on the top left indicates, the fraction of the low-metallicity particles relative to the whole sample is $\sim\,1.2\%$. In the central row, the mean value of azimuthal action $\langle J_{\phi} \rangle$ is shown for each pixel instead of the stellar density. No pseudo star has an initial azimuthal action $J_{\phi}$ larger than $1000\,\kms\,\kpc$, which is reasonable since the pseudo stars are sampled by a truncated DF model. The bottom row shows the low-metallicity particles with $J_{z}<100\,\kms\,\kpc$, which corresponds to a very thin-disc like behaviour. Only very few stars belong to this population, since we choose a large value for $J_{z0}$ in the bulge model.  

The central column of Figure~\ref{fig:const_vmp} shows the distribution of the low-metallicity particles at $\rm{T}=3\,\Gyr$ for the constant bar model. Comparing to the density distribution at $\rm{T}=0\,\Gyr$, the shape of the density contour becomes slightly elliptical in the central region, which is the consequence of the perturbation by the constant rotating bar. However, there is no obvious radial excursion of their orbits compared to their initial conditions. This feature is verified in the central row via colour-coded by $\langle J_{\phi} \rangle$, and shows that there are no particles that possess a $J_{\phi}$ value larger than $1000\,\kms\,\kpc$. The major difference between $\rm{T}=3\,\Gyr$ and the initial condition lies in the bottom row. The number of stars with $J_{z}<100\,\kms\,\kpc$ increases from 393 to 12559. The rightmost column shows their distribution at $\rm{T}=5\,\Gyr$, which resemble that at $\rm{T}=3\,\Gyr$, only with a twist on the orientation of the bar's major axis. The orange dot in this column marks the position of the Sun, and the two dashed curves denote the 4 and 6 $\kpc$ circles away from the Sun. The central black line indicates the direction of the bar's major axis, $\phi\,=\,28^{\circ}$ with respect to the Sun. We notice that the number of low-metallicity particles with $J_{z}<100\,\kms\,\kpc$ increases by a very small amount from $\rm{T}=3\,\Gyr$ to $\rm{T}=5\,\Gyr$, which is explained by the pseudo stars in this simulation reaching near-equilibrium in the bar's rotating frame in less than $2\,\Gyr$ after the bar's onset. 

\begin{figure*}
\centering
	\includegraphics[scale=0.60]{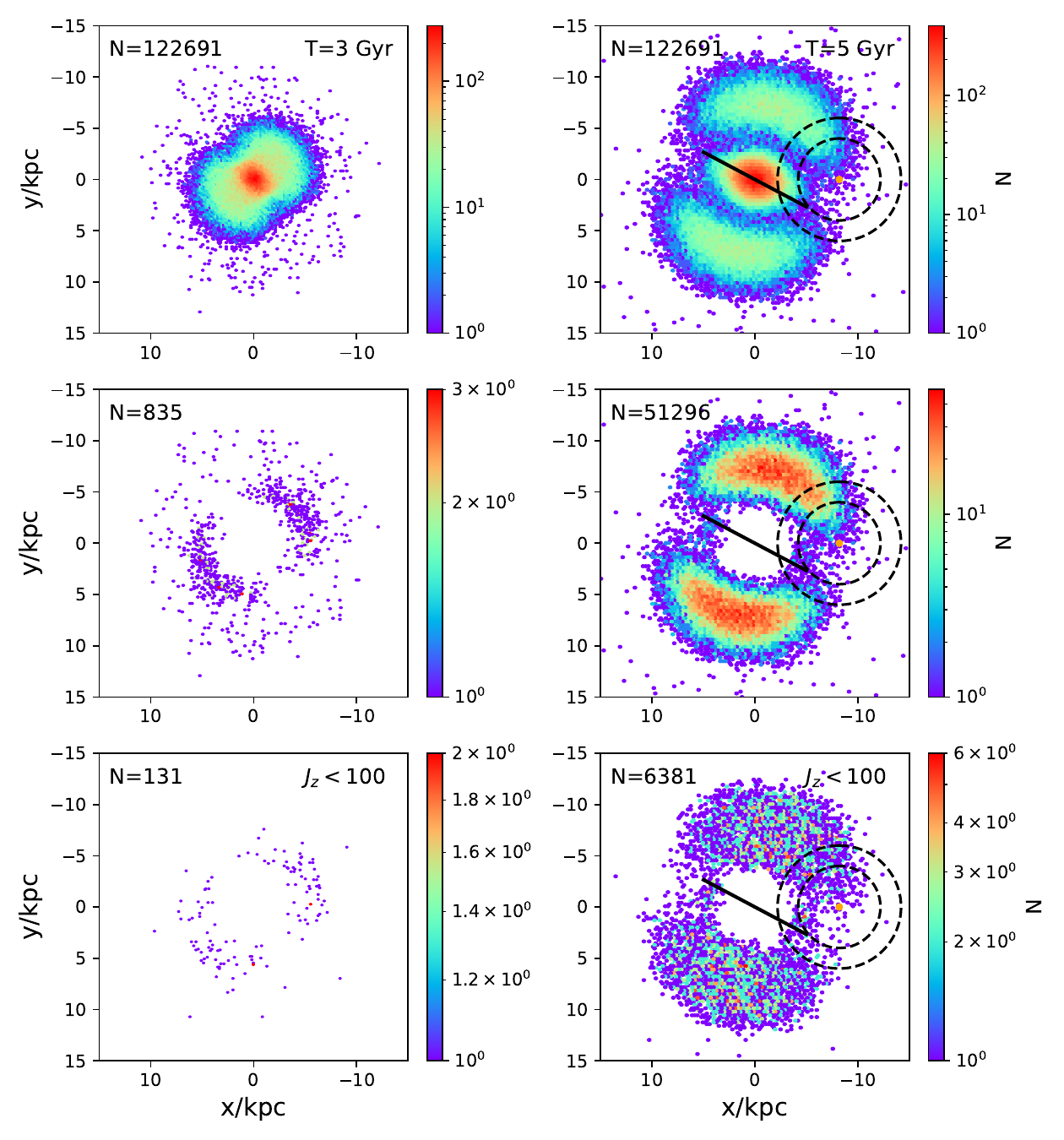}
	\vspace{-0.3cm}
    \caption{The very low-metallicity stars (same as Fig. \ref{fig:const_vmp}) for the boxy/peanut bulge model under the decelerating bar at $\rm{T}=3\,\Gyr$ and $\rm{T}=5\,\Gyr$ respectively. In the central row, we plot the distribution for the low-metallicity particles with $J_{\phi}>1000\,\kms\,\kpc$. The bottom row shows the particles with $J_{\phi}>1000\,\kms\,\kpc\, \& \,J_{z}<100\,\kms\,\kpc$. The orange dot in the right panel marks the position of the Sun, and the two dashed curves denote the distances at 4 and 6 $\kpc$ from the Sun. The pseudo stars in the central region of the Galaxy evolve sharply within the simulation time. Some of these stars are driven outwards by the decelerating bar with the co-rotation trapping. It's clear that at $\rm{T}=5\,\Gyr$, about 6000 pseudo stars show a feature of rotation-dominated orbits with very low metallicities.}
    \label{fig:decel_vmp}
    \vspace{-0.3cm}
\end{figure*}

In the model of a decelerating bar, the left column of Figure~\ref{fig:decel_vmp} shows the distribution of the low-metallicity particles at $\rm{T}=3\,\Gyr$. Unlike the elliptical shape observed in the central region for a constant bar (Fig.~\ref{fig:const_vmp}), we observe the horseshoe orbits under a decelerating bar. This indicates that some orbits are elongated and brought outwards from the central region. In order to prove this, we plot in the central row the distribution of the low-metallicity particles with $J_{\phi}>1000\,\kms\,\kpc$. Only a small portion of them (less than 1 $\%$) are actually showing a disc-like behavior even with Galacto-centric radius $R>10\kpc$. The bottom row shows the particles with $J_{\phi}>1000\,\kms\,\kpc\, \& \,J_{z}<100\,\kms\,\kpc$, which corresponds to thin disc like properties. Contrary to the constant rotating bar, the decelerating bar is still in evolution at $\rm{T}=3\,\Gyr$. 
The low-metallicity particles show a quite different story at $\rm{T}=5\,\Gyr$ and have two clear distinct regions in their distribution in $x-y$ plane (right column). The pseudo stars in the central region are still confined within the bulge but with an elongated shape. On the other hand, two arch shaped regions lying above and below the major axis of the bar show the pseudo stars trapped by the co-rotation resonance of the bar. The central panel shows that 42$\%$ of the low-metallicity stars have $J_{\phi}>1000\,\kms\,\kpc$ at $\rm{T}=5\,\Gyr$, which is more than 60 times the fraction at $\rm{T}=3\,\Gyr$. 5$\%$ of the total number of the low-metallicity stars now have thin disc properties $J_{\phi}>1000\,\kms\,\kpc\, ,J_{z}<100\,\kms\,\kpc$, and are well confined by the co-rotation resonance of the bar. The two dashed circles mark the radius of 4 $\kpc$ and 6 $\kpc$ away from the Sun, radii  within which most of the observed stars with [Fe/H] below $-$2.5 are located. Comparing with the simulations, the fractions of the low-metallicity particles with disc-like behavior located within these two circles are fairly low (shown in the right bottom panel), and amount to about $\sim4\%$ and $\sim10\%$ respectively, relative to the total number $\rm{N}=6381$ in this panel.

The decelerating bar has the ability to trap the most metal poor bulge in the co-rotation resonance region, move them outwards, and empower them with strong rotation as shown in Figure~\ref{fig:decel_vmp}. However, even in this idealized scenario, only a small fraction of the low-metallicity stars from the bulge are currently located in the solar vicinity at $\rm{T}=5\,\Gyr$. Most of the stars are trapped at the co-rotation resonance regions above and below the bar's major axis in the $(x,y)$ plane, which depends on the relative positions between the Sun and the bar. In the simulations, we set the evolution time to be relatively long, because the decelerating bar needs more time to evolve, and some of the features can only be seen several $\Gyr$ after the bar onset. In our experiment, the pattern speed of the bar dropped to a value that is slightly smaller than the current estimated value in our Galaxy \citep{Monari2019,Binney2020,Chiba2021b,Clarke2022,Leung2023,Lucey2023,Gaia2023} at $\rm{T}=5\,\Gyr$. Consequently, the corotation radius stops increasing and the migration of the perturbed stars ceases, we thus set $\rm{T}=5\,\Gyr$ as the end of the simulation.

\begin{figure}
	\includegraphics[width=\columnwidth]{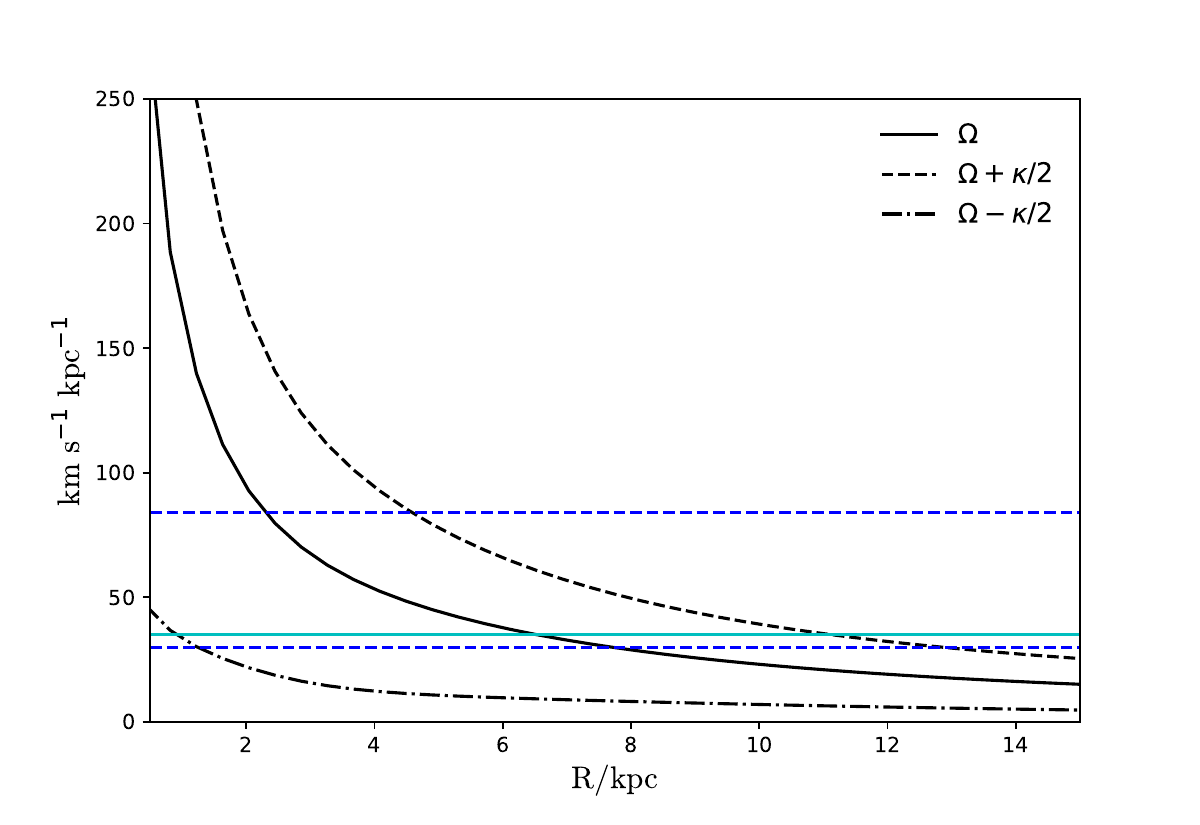}
	\vspace{-0.3cm}
    \caption {Frequency $\Omega\pm\kappa/2$ derived from the background potential as a function of planar radius. The solid cyan line denotes to the bar's pattern speed used in the constant rotating bar model. The co-rotation radius is defined as the planar radius at the intersection between the frequency curve and the pattern speed curve. The dashed blue lines mark the initial and final pattern speed in the decelerating bar model, which shows the change of co-rotation radius during 6 Gyrs. The stars trapped by the bar thus migrate outwards with time until the pattern speed reaches a stable value and the co-rotation radius stops increasing.}
    \label{fig:omega}
    \vspace{-0.3cm}
\end{figure}

To figure out the possible mechanism of the decelerating bar in shaping the bulge distribution shown in Figure~\ref{fig:decel_vmp}, we look into the resonance frequency as a function of planar radius in Figure~\ref{fig:omega}. The solid cyan line corresponds to the bar's pattern speed used in the constant rotating bar model. The two dashed blue lines mark the initial and final pattern speed in the decelerating bar model. The black solid curve is the azimuthal frequency $\Omega$ based on the axisymmetric potential used in this work. Thus the intersections between the bar's pattern speeds and  $\Omega$ denote the location of co-rotation resonance. The intersections of the $\Omega+\kappa/2$ and the bar's pattern speeds denote the OLR (Outer Lindblad resonance) shown as the dashed curve. The similar intersection of $\Omega-\kappa/2$ denotes the ILR (Inner Lindblad resonance) as the dashed-dotted curve. The co-rotation radius of the constant rotating bar is at $R\approx7\kpc$, which is further outside than the sampled particles. The only possible influence of a resonance on the pseudo stars is the ILR. Since the pattern speed does not change in this constant pattern speed model, the resonance trapped radius also stays stable throughout the simulation, which makes stars confined within the initial truncated radius as shown in the upper panel of Figure~\ref{fig:const_vmp}. The case for the decelerating bar is far more interesting. With a larger initial value of the pattern speed, the co-rotation radius for this model is less than 3 $\kpc$ from the centre, well within the sampled pseudo stars' spatial region. The direct consequence is that a fraction of pseudo stars are trapped by the co-rotation resonance since the start of the simulation. As the bar's pattern speed decreases, the co-rotation radius increases to $R\approx8\kpc$ and the number of pseudo stars that are trapped with the co-rotation resonance also increases. Those trapped stars are then brought outwards with the moving of the bar's co-rotation trapped region. 

The discussions on dynamical resonances of the bar in this section are mainly focused on the co-rotation resonance. However, the co-rotation is not the only way that can transfer angular momentum to stars, which can be verified in the lower panel of Figure~\ref{fig:decel_vmp} showing that some pseudo stars are well outside the co-rotation resonance zone of the bar. Another work with similar purposes has been carried out by \citet{Saha2012} through N-body simulations studying the angular momentum transfer from the bar to an initially isotropic non-rotating small classical bulge, bringing stars as far away as the Solar neighbourhodd. They also concluded that the angular momentum transfer happened through the resonances of the bar, but mainly through the ILR. The difference between the two works is that the co-rotation resonance of the bar in \citet{Saha2012} is outside the radius that confine most bulge particles, thus only the ILR plays an important role in transferring the angular momentum to the particles. On the contrary, in our work, the initial co-rotation resonance of the bar is inside the bulge region, leading to more co-rotation resonance trapping of the particles. In our work, the ILR actually also transfers angular momentum to the particles, but with very small contributions. More detailed work should be carried out in the future to study the ability of various bar resonances to transfer angular momentum to the stars in a model like the one presented here.

\subsection{Data and comparisons}{\label{sec:com}}
To compare with the model prediction, we build our sample by cross-matching the $Gaia$ DR3 Radial Velocity Sample \citep[RVS;][]{dr3rvs} with the photometric metallicity catalogues based on the narrow-band $CaHK$ magnitudes from the Pristine survey \citep{martin2023}. There are two catalogues used in this work. One is the Pristine-$Gaia$ synthetic catalogue which uses the $CaHK$ magnitude built from $Gaia$ BP/RP spectro-photometry \citep{xp23}. The other one is the Pristine DR1 based on the real $CaHK$ measurements taken by the Canada-France-Hawaii Telescope, which has a smaller number of stars but higher signal-to-noise compared to the first catalogue. By combining these two samples, we try to include as many low-metallicity RVS stars as possible, and take the metallicity values with smaller uncertainties if stars are duplicated. 

We first select a low-metallicity sample from the giant solutions from Pristine DR1: \texttt{FeH\_Pristine} $>$ $-$4.0 and \texttt{FeH\_Pristine} $\leq$ $-$2.5 and \texttt{mcfrac\_Pristine} $>$ 0.8 and \texttt{FeH\_Pristine\_84th} - \texttt{FeH\_Pristine\_16th} $<$ 0.6. Note that the metallicity column from the Pristine-$Gaia$ synthetic catalogue is \texttt{FeH\_CaHKsyn}. In order to obtain a confident sample of low-metallicity stars, here we use a relatively strict cut in photometric uncertainties ($\delta$[Fe/H]$_{\rm phot}$ $<$ 0.3) and apply the following quality cuts according to the recommended practice from \citet{martin2023}:

\begin{itemize}
    \item \texttt{E(B-V)} $<$ 0.3, relatively strict cut to remove stars with heavy extinctions.
    \item \texttt{Pvar} $\leq$ 0.3, relatively strict cut to remove potentially variable stars.
    \item \texttt{RUWE} $<$ 1.4 from the $Gaia$ quality cuts.
    \item $C^{\ast}$ $<$ 3$\times$ $\sigma_{C^{\ast}}$, relatively loose cut to remove stars deviate more than 3$\sigma$ from the corrected flux excess \citep{edr3phot}.
\end{itemize}

We then filter out the sample using \texttt{parallax$\_$over$\_$error} $\geq$ 5, and manually select stars around the giant branch from the color absolute magnitude diagram. This procedure discards the dwarf stars that get into the sample from the previous selection criteria, which suffer from more severe contamination from the metal-rich population. The final sample has 3259 stars that have uncertainty in distance less than 20$\%$.

\begin{figure}
	\includegraphics[width=\columnwidth]{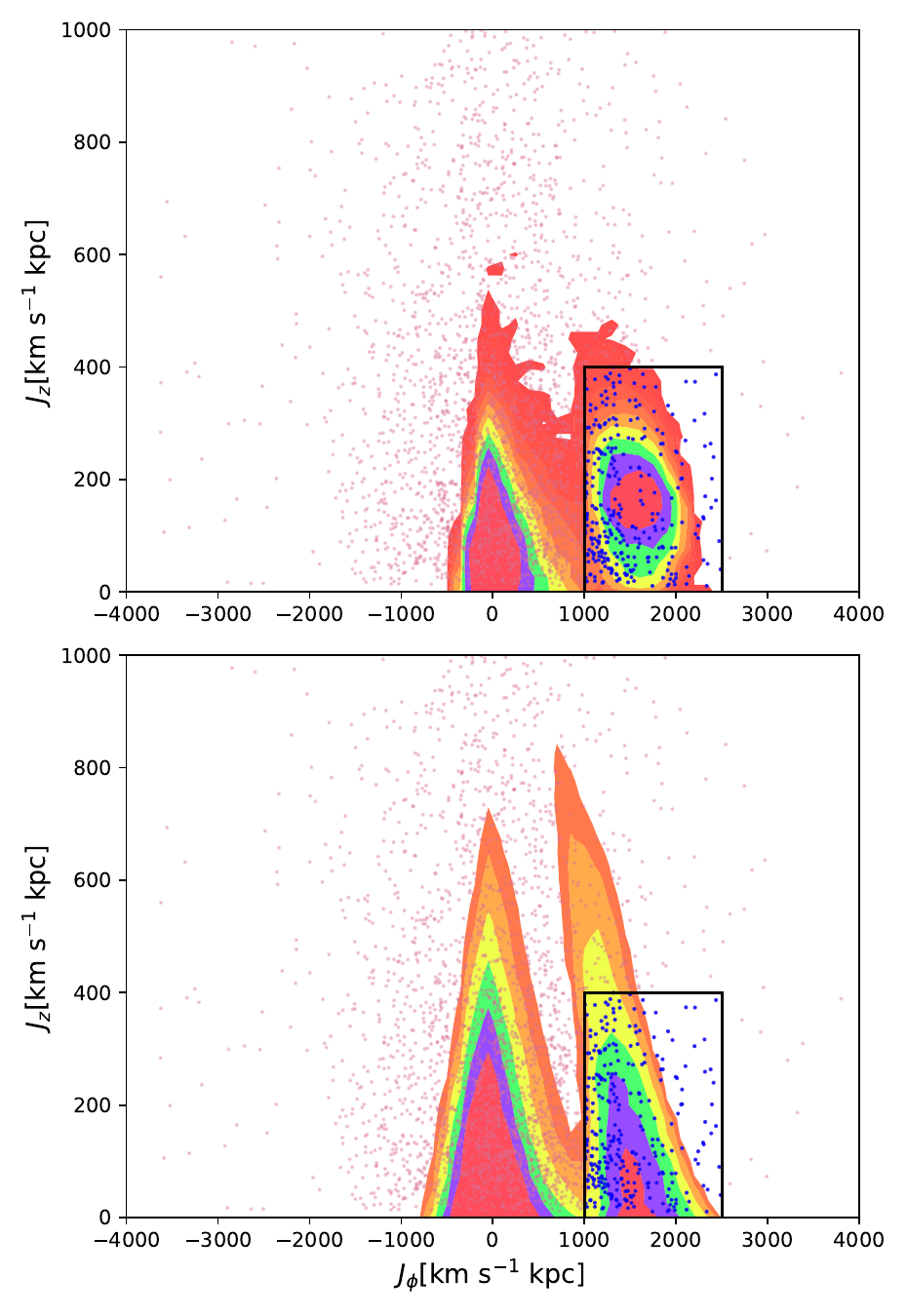}
	\vspace{-0.3cm}
    \caption{In the upper panel, we plot the final $J_{\phi}$-$J_z$ distribution for both the simulation result with a decelerating bar (with a boxy/peanut bulge) and the observational data. The contours show the distribution for very metal poor pseudo stars, with a clear bi-modal distribution. The pseudo stars with $J_{\phi}>1000$ denote the population that is driven outwards by the decelerating bar. The black box represents the selection criteria and the blue stars represent the same stars as those plotted in Figure~\ref{fig:obs}. The simulation result coincides with the observations well. In order to make a comparison with another bulge model, the result from another experiment with a spheroidal bulge is shown in the lower panel. The bi-modality distribution is still clear in this second experiment.}
    \label{fig:sim&data}
    \vspace{-0.3cm}
\end{figure}

A comparison between the observation and simulation in the ($J_{\phi}$, $J_z$) space is shown in Figure~\ref{fig:sim&data}. The color codded contours show the very low-metallicity stars in the simulation with $J_{\phi}$ and $J_z$ values at $\rm{T}=5\Gyr$ in the decelerating bar model. The density increases by a factor of 2 between adjacent contours. We see a clear bimodal distribution where half of the particles keep low $J_{\phi}$ corresponding to those that remain close to the Galactic center (see first row of Fig.\ref{fig:decel_vmp}) and the other half are peaked between $J_{\phi}=1000\sim2000\,\kms\,\kpc$ and are located in the arches further from the center shown in Fig.\ref{fig:decel_vmp}. The final low-metallicity RVS sample is plotted as light purple dots, and the stars in the selection box from \citet{Sestito2020} are highlighted as blue dots. It is clear that the selected rotation-dominated stars overlap with the population with strong rotation from the model. To test the ability of a decelerating bar to bring outwards the low metallicity stars, we also test a model which is introduced in Section~\ref{sec:sph}, where those stars are originally distributed in a spheroid.
The results from a spheroidal bulge also have a clear bi-modalitiy distribution and a second peak that overlaps with the observed stars in the selection box (see the lower panel of Figure~\ref{fig:sim&data}). Note that the final $J_{z}$ distribution from the spheroidal bulge model is more scattered than that from the peanut-shape bulge. These results show that the decelerating bar has the ability to bring the most metal poor stars outwards to thin disc-like orbits regardless of the model used to generate the mock catalogue. It should be noticed that the spheroidal bulge stars do not have chemical information, and thus we plot all the pseudo stars in the lower panel of Figure~\ref{fig:sim&data}.

\begin{figure}
	\includegraphics[width=\columnwidth]{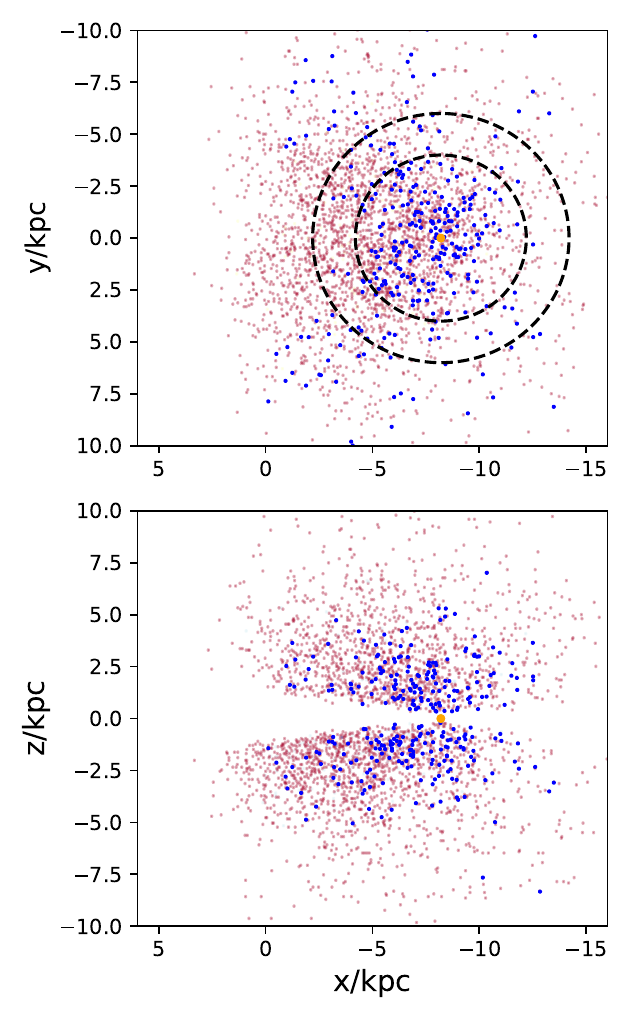}
	\vspace{-0.3cm}
    \caption{In this fig, we plot the space distribution of all the observational data used in this work. The red dot shows all the low-metallicity stars based on $CaHK$ magnitudes, and the blue dot shows the stars with rotation-dominated orbits $1000<J_{\phi}<2500\,\kms\,\kpc\,\&\,J_z<400\,\kms\,\kpc$. The two circles are the same with those in the rightmost panel of Figure~\ref{fig:decel_vmp}.}
    \label{fig:obs}
    \vspace{-0.3cm}
\end{figure}

In Figure~\ref{fig:obs}, we plot the spatial distribution of the final sample shown as light purple dots, including 317 stars (blue dots) selected with $1000<J_{\phi}<2500\,\kms\,\kpc\,\&\,J_z<400\,\kms\,\kpc$ \citep{Sestito2020}. As discussed in Sec. \ref{sec:pred}, the majority of the stars that moved outwards are distributed in the fourth quadrant of the Galaxy and are outside of the solar circles, whereas the observed stars are mostly within. Moreover, the observed stars selected in this work are confined away from the Galactic centre due to the heavy extinction region near the Galactic plane, shown in the $(x,z)$ space in the bottom panel. In summary, the spatial distributions between the model and the observations have fairly small overlaps in space. It is difficult to explain the main origins of these stars observed in the Solar neighborhood through the mechanism of a decelerating bar.

\section{Conclusion}{\label{sec:con}}

In this paper, we carried out test-particle simulations that aim to elucidate the possible origin of the low-metallicity stars with thin disc-like behavior observed recently \citep{Sestito2020,Yuan2023}.

We adopt two different bar models for the simulations. Under the constant pattern speed bar model, the pseudo stars almost keep their initial conditions within the central regions in the Galaxy and have no migration outwards across the disc as time evolves. In addition, the number of stars with $J_{z}<100\,\kms\,\kpc$ increases by a very small amount from $\rm{T}=3\,\Gyr$ to $\rm{T}=5\,\Gyr$, which means that the constant bar reaches a near-equilibrium state earlier than $\rm{T}=2\,\Gyr$ after the bar appears. The scenario of the decelerating bar model is very different. According to the simulation results, the decelerating bar shows the ability to bring the bulge stars outwards by trapping them in the resonance regions. The co-rotation resonance radius of the decelerating bar model is less than 3 $\kpc$ from the centre initially, which allows a fraction of pseudo stars to be trapped by the bar from the start of the emergence of the bar in the simulation. With the decrease of the bar's pattern speed and the increase of the co-rotation radius, the number of pseudo stars that are trapped at the co-rotation resonance increases and these stars also migrate outwards as time evolves. Note that the decelerating bar needs more time to evolve, and the number of stars driven outwards only increases drastically several $\Gyr$ after the bar is created. Possibly the other resonances of the bar also have the ability to transfer angular momentum to stars, such as ILR(see \cite{Saha2012} for example). The more detailed  works should be done in the future to verify the capability of transferring angular momentum to the stars by various resonances of the bar.

We then compare our simulation predictions with the $Gaia$ DR3 RVS sample that has metallicities based on $CaHK$ magnitudes. The results show that the decelerating bar is capable of moving stars from the inner Galaxy outwards on very prograde planar orbits. However, based on the relative positions between the bar and the Sun, only a small fraction of these low-metallicity pseudo bulge stars move to the Solar vicinity: about 10$\%$ of the stars with disc behaviours are located within 6 kpc from us, and only about 4$\%$ within 4 kpc. This possibly indicates that the decelerating bar is unlikely to be the only mechanism for the stars of interests in the Solar neighbourhood, consistent with our results in \citet{Yuan2023}. Most of the migrated stars in our simulation are located in the direction towards the Galactic center where high-quality measurements of photometric metallicities are lacking due to heavy extinction. Note that we concentrated here on the mystery of planar low-metallicity stars, but that the effect of the decelerating bar can similarly (or even more strongly) impact metal-rich stars.

The exact percentages mentioned above can of course depend on the precise way in which the pattern speed of the bar has decelerated. Moreover, our model does not take into account the possibility that short-lived recurring spiral arms with different pattern speeds could have their resonances overlapping with the bar's resonances at different radii \citep{sellwood02, minchev10} and hence enhance the migration of the old stars. Observationally, it had already been noted \citep[e.g.,][]{minchev14, lagarde21} in the past that the oldest stars of the high-$\alpha$ metal poor thick disk tend to have planar orbits with low vertical velocity dispersions compared to slightly younger ones, which had been attributed to a similar migration mechanism.

Moreover, there are a few caveats with the present model. The boxy/peanut bulge can itself be a manifestation resulting from bar instability, meaning that its present-day structure is not necessarily representative of its past one in a growing bar model, which is why it is reassuring that similar results were obtained for a different DF. Besides, the slowing bar model adopted in this paper has not been adjusted to fit the data. The maximum slowing rate is slightly larger than the recent measurement \citep{Chiba2021a} and the final pattern speed of our bar model is smaller than current observational values. These settings enable stars to move over a very large radius range from the very inner Galaxy to our vicinity, which is probably not very realistic. Furthermore, recent studies indicate that the presence of gas may significantly retard or even stop the slowdown of galactic bars \citep{Beane2023}. To further verify the predictions from our simulations, the upcoming WEAVE survey \citep{weave} and the forthcoming 4MOST survey \citep{4most} will fill data in the sparse region towards the Galactic centre. In the long term, the studies of stars in these obscured regions will be greatly benefit from  the Near-Infra-Red (NIR) astrometric survey such as $Gaia$NIR \citep{gaiair}.

\begin{acknowledgements}
CL, AS, BF, GM, GK, and VH acknowledge funding from the ANR grant ANR-20-CE31-0004 (MWDisc). ZY, NFM, BF, GM, and RAI acknowledge funding from the European Research Council (ERC) under the European Unions Horizon 2020 research and innovation programme (grant agreement No. 834148). ZY, NFM, GK, and VH gratefully acknowledge support from the French National Research Agency (ANR) funded project ``Pristine'' (ANR-18-CE31-0017). R.C. thanks support by the the Natural Sciences and Engineering Research Council of Canada (NSERC), [funding reference $\#$DIS-2022- 568580].

This work presents results from the European Space Agency (ESA) space mission Gaia. Gaia data are being processed by the Gaia Data Processing and Analysis Consortium (DPAC). Funding for the DPAC is provided by national institutions, in particular the institutions participating in the Gaia MultiLateral Agreement (MLA). The Gaia mission website is https://www.cosmos.esa.int/gaia. The Gaia archive website is https://archives.esac.esa.int/gaia.

\end{acknowledgements}



\bibliographystyle{aa}
\bibliography{example}

\begin{thebibliography}{64}
\expandafter\ifx\csname natexlab\endcsname\relax\def\natexlab#1{#1}\fi

\bibitem[{{Abdurro'uf} {et~al.}(2022){Abdurro'uf}, {Accetta}, {Aerts}, {Silva
  Aguirre}, {Ahumada}, {Ajgaonkar}, {Filiz Ak}, {Alam}, {Allende Prieto},
  {Almeida}, {Anders}, {Anderson}, {Andrews}, {Anguiano}, {Aquino-Ort{\'\i}z},
  {Arag{\'o}n-Salamanca}, {Argudo-Fern{\'a}ndez}, {Ata}, {Aubert},
  {Avila-Reese}, {Badenes}, {Barb{\'a}}, {Barger}, {Barrera-Ballesteros},
  {Beaton}, {Beers}, {Belfiore}, {Bender}, {Bernardi}, {Bershady}, {Beutler},
  {Bidin}, {Bird}, {Bizyaev}, {Blanc}, {Blanton}, {Boardman}, {Bolton},
  {Boquien}, {Borissova}, {Bovy}, {Brandt}, {Brown}, {Brownstein}, {Brusa},
  {Buchner}, {Bundy}, {Burchett}, {Bureau}, {Burgasser}, {Cabang}, {Campbell},
  {Cappellari}, {Carlberg}, {Wanderley}, {Carrera}, {Cash}, {Chen}, {Chen},
  {Cherinka}, {Chiappini}, {Choi}, {Chojnowski}, {Chung}, {Clerc}, {Cohen},
  {Comerford}, {Comparat}, {da Costa}, {Covey}, {Crane}, {Cruz-Gonzalez},
  {Culhane}, {Cunha}, {Dai}, {Damke}, {Darling}, {Davidson}, {Davies},
  {Dawson}, {De Lee}, {Diamond-Stanic}, {Cano-D{\'\i}az}, {S{\'a}nchez},
  {Donor}, {Duckworth}, {Dwelly}, {Eisenstein}, {Elsworth}, {Emsellem},
  {Eracleous}, {Escoffier}, {Fan}, {Farr}, {Feng}, {Fern{\'a}ndez-Trincado},
  {Feuillet}, {Filipp}, {Fillingham}, {Frinchaboy}, {Fromenteau}, {Galbany},
  {Garc{\'\i}a}, {Garc{\'\i}a-Hern{\'a}ndez}, {Ge}, {Geisler}, {Gelfand},
  {G{\'e}ron}, {Gibson}, {Goddy}, {Godoy-Rivera}, {Grabowski}, {Green},
  {Greener}, {Grier}, {Griffith}, {Guo}, {Guy}, {Hadjara}, {Harding},
  {Hasselquist}, {Hayes}, {Hearty}, {Hern{\'a}ndez}, {Hill}, {Hogg},
  {Holtzman}, {Horta}, {Hsieh}, {Hsu}, {Hsu}, {Huber}, {Huertas-Company},
  {Hutchinson}, {Hwang}, {Ibarra-Medel}, {Chitham}, {Ilha}, {Imig}, {Jaekle},
  {Jayasinghe}, {Ji}, {Johnson}, {Jones}, {J{\"o}nsson}, {Katkov}, {Khalatyan},
  {Kinemuchi}, {Kisku}, {Knapen}, {Kneib}, {Kollmeier}, {Kong}, {Kounkel},
  {Kreckel}, {Krishnarao}, {Lacerna}, {Lane}, {Langgin}, {Lavender}, {Law},
  {Lazarz}, {Leung}, {Leung}, {Lewis}, {Li}, {Li}, {Lian}, {Liang}, {Lin},
  {Lin}, {Lin}, {Lintott}, {Long}, {Longa-Pe{\~n}a}, {L{\'o}pez-Cob{\'a}},
  {Lu}, {Lundgren}, {Luo}, {Mackereth}, {de la Macorra}, {Mahadevan},
  {Majewski}, {Manchado}, {Mandeville}, {Maraston}, {Margalef-Bentabol},
  {Masseron}, {Masters}, {Mathur}, {McDermid}, {Mckay}, {Merloni},
  {Merrifield}, {Meszaros}, {Miglio}, {Di Mille}, {Minniti}, {Minsley},
  {Monachesi}, {Moon}, {Mosser}, {Mulchaey}, {Muna}, {Mu{\~n}oz}, {Myers},
  {Myers}, {Nadathur}, {Nair}, {Nandra}, {Neumann}, {Newman}, {Nidever},
  {Nikakhtar}, {Nitschelm}, {O'Connell}, {Garma-Oehmichen}, {Luan Souza de
  Oliveira}, {Olney}, {Oravetz}, {Ortigoza-Urdaneta}, {Osorio}, {Otter},
  {Pace}, {Padilla}, {Pan}, {Pan}, {Parikh}, {Parker}, {Peirani}, {Pe{\~n}a
  Ram{\'\i}rez}, {Penny}, {Percival}, {Perez-Fournon}, {Pinsonneault},
  {Poidevin}, {Poovelil}, {Price-Whelan}, {B{\'a}rbara de Andrade Queiroz},
  {Raddick}, {Ray}, {Rembold}, {Riddle}, {Riffel}, {Riffel}, {Rix}, {Robin},
  {Rodr{\'\i}guez-Puebla}, {Roman-Lopes}, {Rom{\'a}n-Z{\'u}{\~n}iga}, {Rose},
  {Ross}, {Rossi}, {Rubin}, {Salvato}, {S{\'a}nchez}, {S{\'a}nchez-Gallego},
  {Sanderson}, {Santana Rojas}, {Sarceno}, {Sarmiento}, {Sayres}, {Sazonova},
  {Schaefer}, {Schiavon}, {Schlegel}, {Schneider}, {Schultheis}, {Schwope},
  {Serenelli}, {Serna}, {Shao}, {Shapiro}, {Sharma}, {Shen}, {Shetrone}, {Shu},
  {Simon}, {Skrutskie}, {Smethurst}, {Smith}, {Sobeck}, {Spoo}, {Sprague},
  {Stark}, {Stassun}, {Steinmetz}, {Stello}, {Stone-Martinez},
  {Storchi-Bergmann}, {Stringfellow}, {Stutz}, {Su}, {Taghizadeh-Popp},
  {Talbot}, {Tayar}, {Telles}, {Teske}, {Thakar}, {Theissen}, {Tkachenko},
  {Thomas}, {Tojeiro}, {Hernandez Toledo}, {Troup}, {Trump}, {Trussler},
  {Turner}, {Tuttle}, {Unda-Sanzana}, {V{\'a}zquez-Mata}, {Valentini},
  {Valenzuela}, {Vargas-Gonz{\'a}lez}, {Vargas-Maga{\~n}a}, {Alfaro},
  {Villanova}, {Vincenzo}, {Wake}, {Warfield}, {Washington}, {Weaver},
  {Weijmans}, {Weinberg}, {Weiss}, {Westfall}, {Wild}, {Wilde}, {Wilson},
  {Wilson}, {Wilson}, {Wolf}, {Wood-Vasey}, {Yan}, {Zamora}, {Zasowski},
  {Zhang}, {Zhao}, {Zheng}, {Zheng}, \& {Zhu}}]{Apogee2022}
{Abdurro'uf}, {Accetta}, K., {Aerts}, C., {et~al.} 2022, \apjs, 259, 35

\bibitem[{{Ablimit} {et~al.}(2020){Ablimit}, {Zhao}, {Flynn}, \&
  {Bird}}]{Ablimit2020}
{Ablimit}, I., {Zhao}, G., {Flynn}, C., \& {Bird}, S.~A. 2020, \apjl, 895, L12

\bibitem[{{Arentsen} {et~al.}(2020){Arentsen}, {Starkenburg}, {Martin}, {Hill},
  {Ibata}, {Kunder}, {Schultheis}, {Venn}, {Zucker}, {Aguado}, {Carlberg},
  {Gonz{\'a}lez Hern{\'a}ndez}, {Lardo}, {Longeard}, {Malhan}, {Navarro},
  {S{\'a}nchez-Janssen}, {Sestito}, {Thomas}, {Youakim}, {Lewis}, {Simpson}, \&
  {Wan}}]{Arentsen2020}
{Arentsen}, A., {Starkenburg}, E., {Martin}, N.~F., {et~al.} 2020, \mnras, 491,
  L11

\bibitem[{{Beane} {et~al.}(2023){Beane}, {Hernquist}, {D'Onghia}, {Marinacci},
  {Conroy}, {Qi}, {Sales}, {Torrey}, \& {Vogelsberger}}]{Beane2023}
{Beane}, A., {Hernquist}, L., {D'Onghia}, E., {et~al.} 2023, \apj, 953, 173

\bibitem[{{Binney}(2014)}]{Binney2014}
{Binney}, J. 2014, \mnras, 440, 787

\bibitem[{{Binney}(2020)}]{Binney2020}
{Binney}, J. 2020, \mnras, 495, 895

\bibitem[{{Binney} \& {Tremaine}(2008)}]{Binney2008}
{Binney}, J. \& {Tremaine}, S. 2008, {Galactic Dynamics: Second Edition}

\bibitem[{{Binney} \& {Vasiliev}(2023)}]{Binney2023}
{Binney}, J. \& {Vasiliev}, E. 2023, \mnras, 520, 1832

\bibitem[{{Binney} \& {Vasiliev}(2024)}]{Binney2024}
{Binney}, J. \& {Vasiliev}, E. 2024, \mnras, 527, 1915

\bibitem[{{Bland-Hawthorn} \& {Gerhard}(2016)}]{gerhard16}
{Bland-Hawthorn}, J. \& {Gerhard}, O. 2016, \araa, 54, 529

\bibitem[{{Bovy \& Rix} {et~al.}(2012){Bovy \& Rix}, {Rix}, {Liu}, {Hogg},
  {Beers}, \& {Lee}}]{bovy12b}
{Bovy \& Rix}, J., {Rix}, H.-W., {Liu}, C., {et~al.} 2012, \apj, 753, 148

\bibitem[{{Brook} {et~al.}(2012){Brook}, {Stinson}, {Gibson}, {Kawata},
  {House}, {Miranda}, {Macci{\`o}}, {Pilkington}, {Ro{\v{s}}kar}, {Wadsley}, \&
  {Quinn}}]{brook12}
{Brook}, C.~B., {Stinson}, G.~S., {Gibson}, B.~K., {et~al.} 2012, \mnras, 426,
  690

\bibitem[{{Carollo} {et~al.}(2023){Carollo}, {Christlieb}, {Tissera}, \&
  {Sillero}}]{Carollo2023}
{Carollo}, D., {Christlieb}, N., {Tissera}, P.~B., \& {Sillero}, E. 2023, \apj,
  946, 99

\bibitem[{{Chen} {et~al.}(2023){Chen}, {Hayden}, {Sharma}, {Bland-Hawthorn},
  {Kobayashi}, \& {Karakas}}]{Chen2023}
{Chen}, B., {Hayden}, M.~R., {Sharma}, S., {et~al.} 2023, \mnras, 523, 3791

\bibitem[{{Chiba}(2023)}]{Chiba2023}
{Chiba}, R. 2023, \mnras, 525, 3576

\bibitem[{{Chiba} {et~al.}(2021){Chiba}, {Friske}, \&
  {Sch{\"o}nrich}}]{Chiba2021a}
{Chiba}, R., {Friske}, J. K.~S., \& {Sch{\"o}nrich}, R. 2021, \mnras, 500, 4710

\bibitem[{{Chiba} \& {Sch{\"o}nrich}(2021)}]{Chiba2021b}
{Chiba}, R. \& {Sch{\"o}nrich}, R. 2021, \mnras, 505, 2412

\bibitem[{{Chiba} \& {Sch{\"o}nrich}(2022)}]{Chiba2022}
{Chiba}, R. \& {Sch{\"o}nrich}, R. 2022, \mnras, 513, 768

\bibitem[{{Clarke} \& {Gerhard}(2022)}]{Clarke2022}
{Clarke}, J.~P. \& {Gerhard}, O. 2022, \mnras, 512, 2171

\bibitem[{{Das} \& {Binney}(2016)}]{Das2016a}
{Das}, P. \& {Binney}, J. 2016, \mnras, 460, 1725

\bibitem[{{Das} {et~al.}(2016){Das}, {Williams}, \& {Binney}}]{Das2016b}
{Das}, P., {Williams}, A., \& {Binney}, J. 2016, \mnras, 463, 3169

\bibitem[{{de Jong} {et~al.}(2019){de Jong}, {Agertz}, {Berbel}, {Aird},
  {Alexander}, \& {Amarsi}}]{4most}
{de Jong}, R.~S., {Agertz}, O., {Berbel}, A.~A., {et~al.} 2019, The Messenger,
  175, 3

\bibitem[{{Di Matteo} {et~al.}(2020){Di Matteo}, {Spite}, {Haywood},
  {Bonifacio}, {G{\'o}mez}, {Spite}, \& {Caffau}}]{DiMatteo2020}
{Di Matteo}, P., {Spite}, M., {Haywood}, M., {et~al.} 2020, \aap, 636, A115

\bibitem[{{Dillamore} {et~al.}(2023){Dillamore}, {Belokurov}, {Evans}, \&
  {Davies}}]{Dillamore2023}
{Dillamore}, A.~M., {Belokurov}, V., {Evans}, N.~W., \& {Davies}, E.~Y. 2023,
  \mnras, 524, 3596

\bibitem[{{Dootson} \& {Magorrian}(2022)}]{Dootson2022}
{Dootson}, D. \& {Magorrian}, J. 2022, arXiv e-prints, arXiv:2205.15725

\bibitem[{{El-Badry} {et~al.}(2018){El-Badry}, {Bland-Hawthorn}, {Wetzel},
  {Quataert}, {Weisz}, {Boylan-Kolchin}, {Hopkins}, {Faucher-Gigu{\`e}re},
  {Kere{\v{s}}}, \& {Garrison-Kimmel}}]{el-badry18}
{El-Badry}, K., {Bland-Hawthorn}, J., {Wetzel}, A., {et~al.} 2018, \mnras, 480,
  652

\bibitem[{{Fern{\'a}ndez-Alvar} {et~al.}(2024){Fern{\'a}ndez-Alvar},
  {Kordopatis}, {Hill}, {Battaglia}, {Gallart}, {Gonz{\'a}lez Rivera de la
  Vernhe}, {Thomas}, {Sestito}, {Ardern-Arentsen}, {Martin}, {Viswanathan}, \&
  {Starkenburg}}]{FernandezAlvar2024}
{Fern{\'a}ndez-Alvar}, E., {Kordopatis}, G., {Hill}, V., {et~al.} 2024, arXiv
  e-prints, arXiv:2402.02943

\bibitem[{{Gaia Collaboration} {et~al.}(2023{\natexlab{a}}){Gaia
  Collaboration}, {Drimmel}, {Romero-G{\'o}mez}, {Chemin}, {Ramos}, {Poggio},
  {Ripepi}, {Andrae}, {Blomme}, {Cantat-Gaudin}, {Castro-Ginard}, {Clementini},
  {Figueras}, {Fouesneau}, {Fr{\'e}mat}, {Jardine}, {Khanna}, {Lobel},
  {Marshall}, {Muraveva}, {Brown}, {Vallenari}, {Prusti}, {de Bruijne},
  {Arenou}, {Babusiaux}, {Biermann}, {Creevey}, {Ducourant}, {Evans}, {Eyer},
  {Guerra}, {Hutton}, {Jordi}, {Klioner}, {Lammers}, {Lindegren}, {Luri},
  {Mignard}, {Panem}, {Pourbaix}, {Randich}, {Sartoretti}, {Soubiran}, {Tanga},
  {Walton}, {Bailer-Jones}, {Bastian}, {Jansen}, {Katz}, {Lattanzi}, {van
  Leeuwen}, {Bakker}, {Cacciari}, {Casta{\~n}eda}, {De Angeli}, {Fabricius},
  {Galluccio}, {Guerrier}, {Heiter}, {Masana}, {Messineo}, {Mowlavi},
  {Nicolas}, {Nienartowicz}, {Pailler}, {Panuzzo}, {Riclet}, {Roux},
  {Seabroke}, {Sordo}, {Th{\'e}venin}, {Gracia-Abril}, {Portell}, {Teyssier},
  {Altmann}, {Audard}, {Bellas-Velidis}, {Benson}, {Berthier}, {Burgess},
  {Busonero}, {Busso}, {C{\'a}novas}, {Carry}, {Cellino}, {Cheek}, {Damerdji},
  {Davidson}, {de Teodoro}, {Nu{\~n}ez Campos}, {Delchambre}, {Dell'Oro},
  {Esquej}, {Fern{\'a}ndez-Hern{\'a}ndez}, {Fraile}, {Garabato},
  {Garc{\'\i}a-Lario}, {Gosset}, {Haigron}, {Halbwachs}, {Hambly}, {Harrison},
  {Hern{\'a}ndez}, {Hestroffer}, {Hodgkin}, {Holl}, {Jan{\ss}en}, {Jevardat de
  Fombelle}, {Jordan}, {Krone-Martins}, {Lanzafame}, {L{\"o}ffler}, {Marchal},
  {Marrese}, {Moitinho}, {Muinonen}, {Osborne}, {Pancino}, {Pauwels},
  {Recio-Blanco}, {Reyl{\'e}}, {Riello}, {Rimoldini}, {Roegiers}, {Rybizki},
  {Sarro}, {Siopis}, {Smith}, {Sozzetti}, {Utrilla}, {van Leeuwen}, {Abbas},
  {{\'A}brah{\'a}m}, {Abreu Aramburu}, {Aerts}, {Aguado}, {Ajaj},
  {Aldea-Montero}, {Altavilla}, {{\'A}lvarez}, {Alves}, {Anders}, {Anderson},
  {Anglada Varela}, {Antoja}, {Baines}, {Baker}, {Balaguer-N{\'u}{\~n}ez},
  {Balbinot}, {Balog}, {Barache}, {Barbato}, {Barros}, {Barstow},
  {Bartolom{\'e}}, {Bassilana}, {Bauchet}, {Becciani}, {Bellazzini},
  {Berihuete}, {Bernet}, {Bertone}, {Bianchi}, {Binnenfeld}, {Blanco-Cuaresma},
  {Boch}, {Bombrun}, {Bossini}, {Bouquillon}, {Bragaglia}, {Bramante},
  {Breedt}, {Bressan}, {Brouillet}, {Brugaletta}, {Bucciarelli}, {Burlacu},
  {Butkevich}, {Buzzi}, {Caffau}, {Cancelliere}, {Carballo}, {Carlucci},
  {Carnerero}, {Carrasco}, {Casamiquela}, {Castellani}, {Chaoul}, {Charlot},
  {Chiaramida}, {Chiavassa}, {Chornay}, {Comoretto}, {Contursi}, {Cooper},
  {Cornez}, {Cowell}, {Crifo}, {Cropper}, {Crosta}, {Crowley}, {Dafonte},
  {Dapergolas}, {David}, {de Laverny}, {De Luise}, {De March}, {De Ridder}, {de
  Souza}, {de Torres}, {del Peloso}, {del Pozo}, {Delbo}, {Delgado}, {Delisle},
  {Demouchy}, {Dharmawardena}, {Di Matteo}, {Diakite}, {Diener}, {Distefano},
  {Dolding}, {Enke}, {Fabre}, {Fabrizio}, {Faigler}, {Fedorets}, {Fernique},
  {Fournier}, {Fouron}, {Fragkoudi}, {Gai}, {Garcia-Gutierrez},
  {Garcia-Reinaldos}, {Garc{\'\i}a-Torres}, {Garofalo}, {Gavel}, {Gavras},
  {Gerlach}, {Geyer}, {Giacobbe}, {Gilmore}, {Girona}, {Giuffrida}, {Gomel},
  {Gomez}, {Gonz{\'a}lez-N{\'u}{\~n}ez}, {Gonz{\'a}lez-Santamar{\'\i}a},
  {Gonz{\'a}lez-Vidal}, {Granvik}, {Guillout}, {Guiraud},
  {Guti{\'e}rrez-S{\'a}nchez}, {Guy}, {Hatzidimitriou}, {Hauser}, {Haywood},
  {Helmer}, {Helmi}, {Sarmiento}, {Hidalgo}, {H{\l}adczuk}, {Hobbs}, {Holland},
  {Huckle}, {Jasniewicz}, {Jean-Antoine Piccolo}, {Jim{\'e}nez-Arranz},
  {Juaristi Campillo}, {Julbe}, {Karbevska}, {Kervella}, {Kordopatis}, {Korn},
  {K{\'o}sp{\'a}l}, {Kostrzewa-Rutkowska}, {Kruszy{\'n}ska}, {Kun}, {Laizeau},
  {Lambert}, {Lanza}, {Lasne}, {Le Campion}, {Lebreton}, {Lebzelter}, {Leccia},
  {Leclerc}, {Lecoeur-Taibi}, {Liao}, {Licata}, {Lindstr{\o}m}, {Lister},
  {Livanou}, {Lorca}, {Loup}, {Madrero Pardo}, {Magdaleno Romeo}, {Managau},
  {Mann}, {Manteiga}, {Marchant}, {Marconi}, {Marcos}, {Marcos Santos},
  {Mar{\'\i}n Pina}, {Marinoni}, {Marocco}, {Martin Polo},
  {Mart{\'\i}n-Fleitas}, {Marton}, {Mary}, {Masip}, {Massari},
  {Mastrobuono-Battisti}, {Mazeh}, {McMillan}, {Messina}, {Michalik}, {Millar},
  {Mints}, {Molina}, {Molinaro}, {Moln{\'a}r}, {Monari}, {Mongui{\'o}},
  {Montegriffo}, {Montero}, {Mor}, {Mora}, {Morbidelli}, {Morel}, {Morris},
  {Murphy}, {Musella}, {Nagy}, {Noval}, {Oca{\~n}a}, {Ogden}, {Ordenovic},
  {Osinde}, {Pagani}, {Pagano}, {Palaversa}, {Palicio}, {Pallas-Quintela},
  {Panahi}, {Payne-Wardenaar}, {Pe{\~n}alosa Esteller}, {Penttil{\"a}},
  {Pichon}, {Piersimoni}, {Pineau}, {Plachy}, {Plum}, {Pr{\v{s}}a}, {Pulone},
  {Racero}, {Ragaini}, {Rainer}, {Raiteri}, {Ramos-Lerate}, {Re Fiorentin},
  {Regibo}, {Richards}, {Rios Diaz}, {Riva}, {Rix}, {Rixon}, {Robichon},
  {Robin}, {Robin}, {Roelens}, {Rogues}, {Rohrbasser}, {Rowell}, {Royer}, {Ruz
  Mieres}, {Rybicki}, {Sadowski}, {S{\'a}ez N{\'u}{\~n}ez}, {Sagrist{\`a}
  Sell{\'e}s}, {Sahlmann}, {Salguero}, {Samaras}, {Sanchez Gimenez}, {Sanna},
  {Santove{\~n}a}, {Sarasso}, {Schultheis}, {Sciacca}, {Segol}, {Segovia},
  {S{\'e}gransan}, {Semeux}, {Shahaf}, {Siddiqui}, {Siebert}, {Siltala},
  {Silvelo}, {Slezak}, {Slezak}, {Smart}, {Snaith}, {Solano}, {Solitro},
  {Souami}, {Souchay}, {Spagna}, {Spina}, {Spoto}, {Steele},
  {Steidelm{\"u}ller}, {Stephenson}, {S{\"u}veges}, {Surdej}, {Szabados},
  {Szegedi-Elek}, {Taris}, {Taylor}, {Teixeira}, {Tolomei}, {Tonello}, {Torra},
  {Torra}, {Torralba Elipe}, {Trabucchi}, {Tsounis}, {Turon}, {Ulla}, {Unger},
  {Vaillant}, {van Dillen}, {van Reeven}, {Vanel}, {Vecchiato}, {Viala},
  {Vicente}, {Voutsinas}, {Weiler}, {Wevers}, {Wyrzykowski}, {Yoldas}, {Yvard},
  {Zhao}, {Zorec}, {Zucker}, \& {Zwitter}}]{Gaia2023}
{Gaia Collaboration}, {Drimmel}, R., {Romero-G{\'o}mez}, M., {et~al.}
  2023{\natexlab{a}}, \aap, 674, A37

\bibitem[{{Gaia Collaboration} {et~al.}(2023{\natexlab{b}}){Gaia
  Collaboration}, {Montegriffo}, {Bellazzini}, {De Angeli}, {Andrae},
  {Barstow}, {Bossini}, {Bragaglia}, \& {Burgess}}]{xp23}
{Gaia Collaboration}, {Montegriffo}, P., {Bellazzini}, M., {et~al.}
  2023{\natexlab{b}}, \aap, 674, A33

\bibitem[{{Gallart} {et~al.}(2019){Gallart}, {Bernard}, {Brook}, {Ruiz-Lara},
  {Cassisi}, {Hill}, \& {Monelli}}]{gallart19}
{Gallart}, C., {Bernard}, E.~J., {Brook}, C.~B., {et~al.} 2019, Nature
  Astronomy, 3, 932

\bibitem[{{Hobbs} {et~al.}(2016){Hobbs}, {H{\o}g}, {Mora}, {Crowley},
  {McMillan}, {Ranalli}, {Heiter}, {Jordi}, {Hambly}, {Church}, {Anthony},
  {Tanga}, {Chemin}, {Portell}, {Jim{\'e}nez-Esteban}, {Klioner}, {Mignard},
  {Fynbo}, {Wyrzykowski}, {Rybicki}, {Anderson}, {Cellino}, {Fabricius},
  {Davidson}, \& {Lindegren}}]{gaiair}
{Hobbs}, D., {H{\o}g}, E., {Mora}, A., {et~al.} 2016, arXiv e-prints,
  arXiv:1609.07325

\bibitem[{{Jin} {et~al.}(2023){Jin}, {Trager}, {Dalton}, {Aguerri}, \&
  {Drew}}]{weave}
{Jin}, S., {Trager}, S.~C., {Dalton}, G.~B., {Aguerri}, J. A.~L., \& {Drew},
  J.~E. 2023, \mnras [\eprint[arXiv]{2212.03981}]

\bibitem[{{Katz} {et~al.}(2023){Katz}, {Sartoretti}, {Guerrier}, {Panuzzo},
  {Seabroke}, \& {Th{\'e}venin}}]{dr3rvs}
{Katz}, D., {Sartoretti}, P., {Guerrier}, A., {et~al.} 2023, \aap, 674, A5

\bibitem[{{Lagarde} {et~al.}(2021){Lagarde}, {Reyl{\'e}}, {Chiappini}, {Mor},
  {Anders}, {Figueras}, {Miglio}, {Romero-G{\'o}mez}, {Antoja}, {Cabral},
  {Salomon}, {Robin}, {Bienaym{\'e}}, {Soubiran}, {Cornu}, \&
  {Montillaud}}]{lagarde21}
{Lagarde}, N., {Reyl{\'e}}, C., {Chiappini}, C., {et~al.} 2021, \aap, 654, A13

\bibitem[{{Leung} {et~al.}(2023){Leung}, {Bovy}, {Mackereth}, {Hunt}, {Lane},
  \& {Wilson}}]{Leung2023}
{Leung}, H.~W., {Bovy}, J., {Mackereth}, J.~T., {et~al.} 2023, \mnras, 519, 948

\bibitem[{{Li} \& {Binney}(2022{\natexlab{a}})}]{Li2022a}
{Li}, C. \& {Binney}, J. 2022{\natexlab{a}}, \mnras, 510, 4706

\bibitem[{{Li} \& {Binney}(2022{\natexlab{b}})}]{Li2022b}
{Li}, C. \& {Binney}, J. 2022{\natexlab{b}}, \mnras, 516, 3454

\bibitem[{{Li} {et~al.}(2023){Li}, {Siebert}, {Monari}, {Famaey}, \&
  {Rozier}}]{Li2023}
{Li}, C., {Siebert}, A., {Monari}, G., {Famaey}, B., \& {Rozier}, S. 2023,
  \mnras, 524, 6331

\bibitem[{{Lucey} {et~al.}(2023){Lucey}, {Pearson}, {Hunt}, {Hawkins}, {Ness},
  {Petersen}, {Price-Whelan}, \& {Weinberg}}]{Lucey2023}
{Lucey}, M., {Pearson}, S., {Hunt}, J. A.~S., {et~al.} 2023, \mnras, 520, 4779

\bibitem[{{Martin} {et~al.}(2023){Martin}, {Starkenburg}, {Yuan}, {Fouesneau},
  {Arentsen}, \& {De Angeli}}]{martin2023}
{Martin}, N.~F., {Starkenburg}, E., {Yuan}, Z., {et~al.} 2023, arXiv e-prints,
  arXiv:2308.01344

\bibitem[{{Minchev} {et~al.}(2014){Minchev}, {Chiappini}, {Martig},
  {Steinmetz}, {de Jong}, {Boeche}, {Scannapieco}, {Zwitter}, {Wyse}, {Binney},
  {Bland-Hawthorn}, {Bienaym{\'e}}, {Famaey}, {Freeman}, {Gibson}, {Grebel},
  {Gilmore}, {Helmi}, {Kordopatis}, {Lee}, {Munari}, {Navarro}, {Parker},
  {Quillen}, {Reid}, {Siebert}, {Siviero}, {Seabroke}, {Watson}, \&
  {Williams}}]{minchev14}
{Minchev}, I., {Chiappini}, C., {Martig}, M., {et~al.} 2014, \apjl, 781, L20

\bibitem[{{Minchev} \& {Famaey}(2010)}]{minchev10}
{Minchev}, I. \& {Famaey}, B. 2010, \apj, 722, 112

\bibitem[{{Monari} {et~al.}(2019){Monari}, {Famaey}, {Siebert}, {Wegg}, \&
  {Gerhard}}]{Monari2019}
{Monari}, G., {Famaey}, B., {Siebert}, A., {Wegg}, C., \& {Gerhard}, O. 2019,
  \aap, 626, A41

\bibitem[{{Portail} {et~al.}(2017){Portail}, {Gerhard}, {Wegg}, \&
  {Ness}}]{Portail2017}
{Portail}, M., {Gerhard}, O., {Wegg}, C., \& {Ness}, M. 2017, \mnras, 465, 1621

\bibitem[{{Riello} {et~al.}(2021){Riello}, {De Angeli}, \& {Evans}}]{edr3phot}
{Riello}, M., {De Angeli}, F., \& {Evans}. 2021, \aap, 649, A3

\bibitem[{{Saha} {et~al.}(2012){Saha}, {Martinez-Valpuesta}, \&
  {Gerhard}}]{Saha2012}
{Saha}, K., {Martinez-Valpuesta}, I., \& {Gerhard}, O. 2012, \mnras, 421, 333

\bibitem[{{Sanders} \& {Binney}(2015)}]{Sanders2015}
{Sanders}, J.~L. \& {Binney}, J. 2015, \mnras, 449, 3479

\bibitem[{{Sch{\"o}nrich} \& {Binney}(2009)}]{Schonrich2009}
{Sch{\"o}nrich}, R. \& {Binney}, J. 2009, \mnras, 396, 203

\bibitem[{{Sellwood}(2014)}]{Sellwood2014}
{Sellwood}, J.~A. 2014, Reviews of Modern Physics, 86, 1

\bibitem[{{Sellwood} \& {Binney}(2002)}]{sellwood02}
{Sellwood}, J.~A. \& {Binney}, J.~J. 2002, \mnras, 336, 785

\bibitem[{{Sellwood} \& {Wilkinson}(1993)}]{Sellwood1993}
{Sellwood}, J.~A. \& {Wilkinson}, A. 1993, Reports on Progress in Physics, 56,
  173

\bibitem[{{Sestito} {et~al.}(2021){Sestito}, {Buck}, {Starkenburg}, {Martin},
  {Navarro}, {Venn}, {Obreja}, {Jablonka}, \& {Macci{\`o}}}]{Sestito2021}
{Sestito}, F., {Buck}, T., {Starkenburg}, E., {et~al.} 2021, \mnras, 500, 3750

\bibitem[{{Sestito} {et~al.}(2020){Sestito}, {Martin}, {Starkenburg},
  {Arentsen}, {Ibata}, {Longeard}, {Kielty}, {Youakim}, {Venn}, {Aguado},
  {Carlberg}, {Gonz{\'a}lez Hern{\'a}ndez}, {Hill}, {Jablonka}, {Kordopatis},
  {Malhan}, {Navarro}, {S{\'a}nchez-Janssen}, {Thomas}, {Tolstoy}, {Wilson},
  {Palicio}, {Bialek}, {Garcia-Dias}, {Lucchesi}, {North}, {Osorio}, {Patrick},
  \& {Peralta de Arriba}}]{Sestito2020}
{Sestito}, F., {Martin}, N.~F., {Starkenburg}, E., {et~al.} 2020, \mnras, 497,
  L7

\bibitem[{{Sharma} {et~al.}(2021){Sharma}, {Hayden}, \&
  {Bland-Hawthorn}}]{Sharma2021}
{Sharma}, S., {Hayden}, M.~R., \& {Bland-Hawthorn}, J. 2021, \mnras, 507, 5882

\bibitem[{{Shen} {et~al.}(2010){Shen}, {Rich}, {Kormendy}, {Howard}, {De
  Propris}, \& {Kunder}}]{Shen2010}
{Shen}, J., {Rich}, R.~M., {Kormendy}, J., {et~al.} 2010, \apjl, 720, L72

\bibitem[{{Sormani} {et~al.}(2022){Sormani}, {Gerhard}, {Portail}, {Vasiliev},
  \& {Clarke}}]{Sormani2022}
{Sormani}, M.~C., {Gerhard}, O., {Portail}, M., {Vasiliev}, E., \& {Clarke}, J.
  2022, \mnras, 514, L1

\bibitem[{{Starkenburg} {et~al.}(2017){Starkenburg}, {Oman}, {Navarro},
  {Crain}, {Fattahi}, {Frenk}, {Sawala}, \& {Schaye}}]{starkenburg17}
{Starkenburg}, E., {Oman}, K.~A., {Navarro}, J.~F., {et~al.} 2017, \mnras, 465,
  2212

\bibitem[{{Tremaine} \& {Weinberg}(1984)}]{Tremaine1984}
{Tremaine}, S. \& {Weinberg}, M.~D. 1984, \mnras, 209, 729

\bibitem[{{Vasiliev}(2019)}]{Vasiliev2019}
{Vasiliev}, E. 2019, \mnras, 482, 1525

\bibitem[{{Wegg} {et~al.}(2019){Wegg}, {Gerhard}, \& {Bieth}}]{Wegg2019}
{Wegg}, C., {Gerhard}, O., \& {Bieth}, M. 2019, \mnras, 485, 3296

\bibitem[{{Wegg} {et~al.}(2015){Wegg}, {Gerhard}, \& {Portail}}]{Wegg2015}
{Wegg}, C., {Gerhard}, O., \& {Portail}, M. 2015, \mnras, 450, 4050

\bibitem[{{Weinberg}(1989)}]{Weinberg1989}
{Weinberg}, M.~D. 1989, \mnras, 239, 549

\bibitem[{{Xiang} \& {Rix}(2022)}]{xiang22}
{Xiang}, M. \& {Rix}, H.-W. 2022, \nat, 603, 599

\bibitem[{{Yuan} {et~al.}(2023){Yuan}, {Li}, {Martin}, {Monari}, {Famaey},
  {Siebert}, {Ardern-Arentsen}, {Sestito}, {Thomas}, {Hill}, {Ibata},
  {Kordopatis}, {Starkenburg}, \& {Viswanathan}}]{Yuan2023}
{Yuan}, Z., {Li}, C., {Martin}, N.~F., {et~al.} 2023, arXiv e-prints,
  arXiv:2311.08464

\end{thebibliography}



\appendix
\section{Speed parameter}\label{sec:spp}

The response of stars to a slowing bar depends crucially on the bar's slowing rate \citep{Tremaine1984,Chiba2021a}. From a dynamical viewpoint, whether the bar is slowing down 'slowly' or 'rapidly' is classified by the dimensionless speed parameter $s$ \citep{Tremaine1984}, which is the ratio of the characteristic libration period to the time the resonance takes to move by its width \citep{Chiba2023}. When $0<s<1$, the bar's resonance moves slowly such that a finite volume of phase space surrounding the resonance is trapped (the \textit{slow regime}). Beyond $s=1$, the evolution is too fast for any star to remain trapped in resonance (the \textit{fast regime}).

\begin{figure}
	\includegraphics[width=\columnwidth]{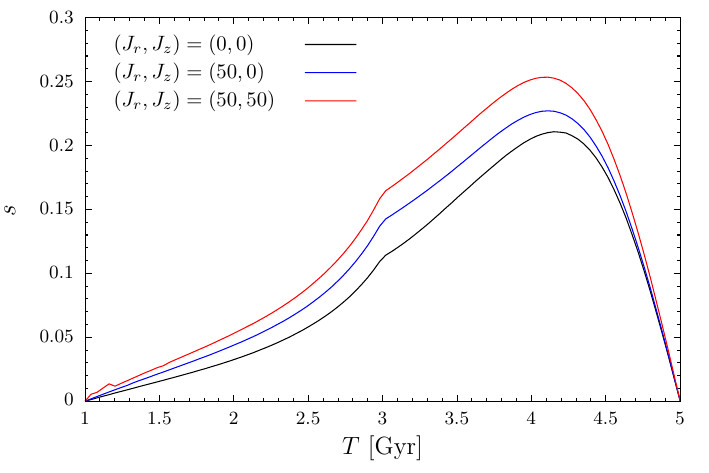}
	\vspace{-0.3cm}
    \caption{The speed parameter of our slowing bar model for the corotation resonance. The black, blue, and red curves denote the computation for $(J_r,J_z)=(0,0),(50,0),(50,50) \kms \kpc$ respectively.}
    \label{fig:s}
    \vspace{-0.3cm}
\end{figure}

Figure~\ref{fig:s} shows the speed parameter of our slowing bar model for the corotation resonance. We calculated $s$ for three different radial and vertical actions $(J_r,J_z)=(0,0),(50,0),(50,50) \kms \kpc$ , which are integral of motions at the corotation resonance (the fast actions). For all three actions, the speed parameter is below unity, indicating that our slowing bar model resonantly traps and drags a sizeable amount of stars at corotation. The speed parameter initially rises and later decays, following the variation in the bar's slowing rate $\dot{\Omega}_\textrm{p}(t)$ (cf. Fig.~\ref{fig:wp}). The bump at $T=3Gyr$ is inherited from the behavior of $\dot{\Omega}_\textrm{p}$.

\end{document}